\documentclass[twocolumn]{revtex4}
\usepackage{amsfonts}
\usepackage{amsmath}
\usepackage{amssymb}
\usepackage{ragged2e}
\usepackage{graphicx}
\usepackage{bm}
\usepackage{rotating}
\usepackage[usenames,dvipsnames]{color}

\definecolor{darkblue}{RGB}{0,0,196}

\begin{document}

\title{\large An energy independent scaling of transverse momentum spectra of
direct (prompt) photons from two-body processes in high-energy
proton-proton collisions \vspace{0.25cm}}

\author{Qiang~Zhang$^{1,}$\footnote{qiangzhangsx@163.com;
2504279475@qq.com}, Ya-Qin Gao$^{2,}$\footnote{gyq610@163.com;
gaoyaqin@tyust.edu.cn}, Fu-Hu
Liu$^{1,2,}$\footnote{Correspondence: fuhuliu@163.com;
fuhuliu@sxu.edu.cn}, Khusniddin K.
Olimov$^{3,}$\footnote{Correspondence: khkolimov@gmail.com;
kh.olimov@uzsci.net} \vspace{0.25cm}}

\affiliation{$^1$Institute of Theoretical Physics, State Key
Laboratory of Quantum Optics and Quantum Optics Devices \&
Collaborative Innovation Center of Extreme Optics, Shanxi
University, Taiyuan 030006, China
\\
$^2$Department of Physics, Taiyuan University of Science and
Technology, Taiyuan 030024, China
\\
$^3$Laboratory of High Energy Physics, Physical-Technical
Institute of Uzbekistan Academy of Sciences, Chingiz Aytmatov str.
$2^b$, Tashkent 100084, Uzbekistan}

\begin{abstract}

\vspace{0.25cm}

\noindent {\bf Abstract:} Transverse momentum spectra of direct
(prompt) photons from two-body processes in high-energy
proton-proton (p+p) collisions are analyzed in this paper. We
collected the experimental invariant cross-sections at
mid-rapidity in p+p collisions measured by the UA6, CCOR, R806,
R110, PHENIX, NA24, CMS, ALICE, and ATLAS Collaborations over a
center-of-mass energy range from 24.3 GeV to 13 TeV. In fitting
the data, we used different kinds of functions which include the
revised Tsallis--Pareto-type function (the TP-like function) at
the particle level, the convolution of two TP-like functions at
the quark level, and the root-sum-of-squares of two revised
Tsallis-like functions in which the quark chemical potentials $\mu
_i=\mu_B/3$ or $\mu_i=0$, where $\mu_B$ is the baryon chemical
potential. We have extracted the values of three main free
parameters: the effective temperature $T$, power index $n_0$ (or
entropy index $q$), and correction index $a_0$. After analyzing
the changing trends of the parameters, we found that $T$, $q$, and
$a_0$ increase and $n_0$ decreases with the increase of collision
energy. Based on the analyses of transverse momentum spectra, an
energy independent scaling, i.e. the $x_T$ scaling, is obtained.
\\
\\
{\bf Keywords:} Transverse momentum spectra, direct (prompt)
photon, TP-like function, revised Tsallis-like function, $x_T$
scaling
\\
\\

\end{abstract}

\maketitle

\section{Introduction}

Quantum Chromodynamics (QCD) and similar strong interaction
theories predict that Quark-Gluon Plasma (QGP) may be generated in
a high-temperature and high-density environment within a few
microseconds after the little Big Bang in heavy ion collisions at
high energy~\cite{1,2,3,4,5}. Understanding the dynamic evolution
of quarks and gluons in nuclei is an important issue for modern
nuclear physics~\cite{2,3,4,5}. The main reason for studying
high-energy heavy ion collisions is to understand the formation
and properties of QGP~\cite{6,15,16,17}. High-energy proton-proton
(p+p), proton-nucleus, and nucleus-nucleus collisions can provide
a high-temperature and/or high-density environment, which is
similar to the early days of the Big Bang in the
universe~\cite{7,8,9,10}. In the evolution process of high-energy
collisions~\cite{11,12,13}, the chemical and kinetic freeze-outs
are two very important stages. In the chemical freeze-out stage,
the types and ratios of particles are fixed, and the system is in
chemical equilibrium. In the kinetic freeze-out stage, the elastic
collisions of particles disappear, and, hence, the kinematic
properties of particles remain unchanged afterwards~\cite{12,14}.

During the whole process of high-energy collisions, the particles
are continuously emitted outwards. The types and ratios of
different particles are changeable before the kinetic free-out
stage. In the collisions, some final products or particles are
produced from initial and/or intermediate stages. When some
primary particles produced in collisions are difficult to be
detected and measured directly, measuring the final particles that
carry a large amount information on system evolution becomes an
important choice and tool for researchers. One can extract the
information on collision process by analyzing the spectra of final
particles. The information about evolution and various properties
of the system is deduced by studying the properties of final
particles. This study is useful to understand the evolution of
early universe and structure of tight stars, the deconfinement of
quarks and formation of QGP, as well as the mechanisms of particle
production~\cite{15,16,17}. In abundant experimental data, the
transverse momentum ($p_T$) spectra of particles are particularly
important~\cite{18} due to the fact that they reflect the
excitation degree and transverse structure of the system. Indeed,
this information is related to the QGP and particle
production~\cite{18a,18b,18c,18d,18e,18f,18g,18h}, because QGP
eventually decomposes into final particles.

Direct photons are defined as all photons generated directly in
the scattering process, but not from hadron decay. Direct photons
provide a powerful tool for studying QGP. As a part of direct
photons, prompt photons are defined as those coming from hard
scattering of two partons from the incident hadrons, or those from
the collinear fragmentation of high-$p_T$ final state
partons~\cite{19,20}. Prompt photons provide a method and approach
for the measurement of perturbation QCD (pQCD), and are also a
probe for the study of the initial state of protons or nuclei. At
the lowest order of pQCD, parton produced photons come from two
hard scattering sub-processes: (i) The quark-gluon Compton
scattering, $qg \to q\gamma$ (main), where $q$, $g$, and $\gamma$
denote quark, gluon, and photon, respectively; (ii) The
quark-antiquark annihilation, $q\bar q \to g\gamma$ (primary),
where $\bar q$ denotes antiquark. In addition, prompt photons can
also be produced by higher-order processes such as bremsstrahlung
or fragmentation. The collinear part of this process has proved to
contribute at the lowest order. We do not need to select the
prompt photons from the direct photons particularly. The study of
$p_T$ spectra of direct (prompt) photons will help to study the
QCD and pQCD processes, and understand some properties of the
initial state of protons or nuclei~\cite{21}.

The invariant or differential cross-sections of direct (prompt)
photons have been extensively studied in high-energy collisions.
Several collaborations at the Large Hadron Collider (LHC) of the
European Organization for Nuclear Research (CERN, abbreviated from
the previous French name of ``Conseil Europ{\'e}enn pour la
Recherche Nucl{\'e}aire"), the Relativistic Heavy Ion Collider
(RHIC) of the Brookhaven National Laboratory (BNL), and other
colliders have measured the $p_T$ spectra of generated direct
(prompt) photons in various collisions. Compared to nuclear
collisions, p+p collisions are the basic and cleaner process with
absences of the influence of nucleon-nucleon correlation and cold
spectator nucleons. Understanding the characteristics of particle
productions is very necessary for researchers to study the
evolution of collision system and interactions among various
particles~\cite{21a,21b,21c,21d,21e,21f,21g,21h}.

The $p_T$ spectrum of direct (prompt) photons is very wide, which
distributes in a range from 0 to above 1000 GeV/$c$ in some cases.
In different $p_T$ regions, the forms of spectrum may be
different~\cite{22}. Generally speaking, the spectrum in low-$p_T$
(intermediate-$p_T$) region is related to the soft (intermediate)
scattering process, and the spectrum in high-$p_T$ region is
related to the hard scattering process in which the direct photons
are also called the prompt photons. However, there are not clear
demarcation points to distinguish various $p_T$ regions. In the
production of multiple particles~\cite{23}, in the case of
considering a given collision energy and the constant particle
species, we usually choose a fitting function with the best effect
for $p_T$ spectrum to extract some quantities such as the rapidity
($y$) or pseudorapidity ($\eta$) density ($dN/dy$ or $dN/d\eta$),
kinetic freeze-out temperature ($T_{kin}$ or $T_0$), average
transverse velocity ($\left\langle \beta_T \right\rangle$ or
$\beta_T$), and etc., where $N$ denotes the number of particles.
In the two-body processes such as $qg\rightarrow q\gamma$ and
$q\bar q\rightarrow g\gamma$, some parameters such as $T_0$ and
$\beta_T$ may be unavailable. Instead, we may use the effective
temperature ($T$) and others.

To fit the $p_T$ spectrum of direct (prompt) photons, different
models or functions can be used. These models or functions
include, but are not limited to, the inverse power-law or Hagedorn
function~\cite{24}, Tsallis--L{\'e}vy~\cite{25,26} or
Tsallis--Pareto-type function~\cite{26,27,28,29}, Bose--Einstein
or Fermi--Dirac distribution, Boltzmann distribution~\cite{30,31},
and so on. After constant exploration, we know that it is
difficult to fit well the spectrum in both the low- and high-$p_T$
regions with only a simple probability density function~\cite{23}.
Therefore, for a wider $p_T$ spectrum, a multi-component (at least
a two-component) function is generally used in the fit~\cite{32}.
In the two-component function, the first component represents the
low-$p_T$ region, while the second component represents the
high-$p_T$ region. However, it is not ideal if the two-component
function is superimposed in terms of usual function (in which the
parameters are correlated) or step function (in which the
connection point is unsmooth)~\cite{33,34}. We hope to find a new
function which can fit well the spectrum of whole $p_T$ region.

In view of the problem of two-component function, in this paper,
we used a few one-component functions to describe the $p_T$
spectrum of direct (prompt) photons and compare their fitting
qualities. We collected the $p_T$ spectra of direct (prompt)
photons from two-body processes in p+p collisions over a
center-of-mass energy ($\sqrt{s}$) range from 24.3 GeV to 13 TeV,
which are measured by a few international collaborations. The
values of parameters are obtained and the trends of parameters are
analyzed. Thus, an energy independent scaling, the $x_T$
scaling~\cite{36,37}, of $p_T$ spectra are obtained and analyzed.

The theoretical formalism used in this paper is based on the
statistical methods (non-extensive statistics). Several models
were tested, like the revised Tsallis--Pareto-type (TP-like)
function at the particle level and the convolution of two TP-like
functions at the parton level as well as a revised Tsallis-like
function at the particle level and the root-sum-of-squares of two
revised Tsallis-like functions at the parton level. This paper
provides useful information for high energy physics
phenomenologists and researches working on non-extensive
statistics applied to scattering processes for addressing direct
(prompt) photon production. The formalism considered in this paper
is widely used to describe the charged particle production in high
energy collisions, and for the first time it is used for the
direct (prompt) photon production by us.

The rest of this paper is structured as follows. The formalism and
method are briefly introduced in Section II. The results and
discussion are given in Section III. In Section IV, we summarize
our main observations and conclusions.

\section{Formalism and method}

\subsection{The TP-like function and the convolution of two functions}

According to refs.~\cite{26,27,28,29}, we know that the
Tsallis--Pareto-type function can be written as
\begin{align}
f(p_T) =& \,\,\frac{A(n_0-1)(n_0-2)}{n_0T[n_0T+m_0(n_0-2)]}p_Tm_T \nonumber\\
&\times\left(1+\frac{m_T-m_0}{n_0T}\right)^{-n_0},
\end{align}
where the effective temperature $T$ and power index $n_0$ are free
parameters, $A$ is the normalization constant, $m_0$ is the rest
mass, and $m_T=\sqrt{p_T^2+m_0^2}$ is the transverse mass.
Equation (1) can be simplified and rewritten to the following
form~\cite{38,39,40,41}:
\begin{align}
f(p_T)=Cp_Tm_T\left(1+\frac{m_T-m_0}{n_0T}\right)^{-n_0},
\end{align}
where $C$ is the normalization constant related to $T$, $n_0$, and
$m_0$.

The above probability density function is expected to fit
simultaneously the spectrum in low- and high-$p_T$ regions.
However, our studies show that the spectrum in very-low-$p_T$
region cannot be fitted well by the equation. Empirically, a
revised form,
\begin{align}
f(p_T)=C(p_Tm_T)^{a_0}\left(1+\frac{m_T-m_0}{n_0T}\right)^{-n_0},
\end{align}
is much better in the fitting treatment, where the power exponent
$a_0$ is a correction factor. The normalization constants in Eqs.
(2) and (3) are different, though both of them are denoted by the
same symbol $C$. Meanwhile, although $a_0$ is dimensionless, $C$
has a dimension so that the units in the two sides of the equation
are the same [$({\rm GeV}/c)^{-1}$ in general]. In addition, we
have used the same symbols, $f(p_T)$, in Eqs. (2) and (3), though
the functions are different.

Equation (3) is in fact a revised Tsallis--Pareto-type function.
For convenience, we call Eq. (3) the TP-like function. The
parameter $T$ used above is not a real temperature, but the
effective temperature. The parameter $n_0$ reflects the degree of
equilibrium of the system in terms of the grand canonical ensemble
for a large amount of events which are indeed true in experiments.
A larger $n_0$ corresponds to a more equilibrium of the system. As
a function that describes the $p_T$ spectrum at the particle
level, Eq. (3) is still not ideal in some cases. Empirically, we
need to consider the convolution of two TP-like
functions~\cite{23}, if we assume that two participant quarks or
partons contribute to the $p_T$ spectrum in terms of two energy
sources.

The first and second energy sources contribute $p_{t1}$ and
$p_{t2}$ to $p_T$ respectively, i.e. $p_T=p_{t1}+p_{t2}$, where
$p_{t1}$ and $p_{t2}$ obey the TP-like function. Due to Eq. (3),
we have the probability density function obeyed by the $i$-th
energy source to be
\begin{align}
f_i(p_{ti})=C_i(p_{ti}m_{ti})^{a_0}
\left(1+\frac{m_{ti}-m_{0i}}{n_0T}\right)^{-n_0},
\end{align}
where $C_i$ is the normalization constant related to the
parameters, $m_{ti}=\sqrt{p_{ti}^2+m_{0i}^2}$ is the transverse
mass, and $m_{0i}$ is the constituent mass for the $i$-th quark or
parton. In the two-body process, the lightest constituent mass
(0.31 GeV/$c^2$) of quark $u$ or $d$ is used for each participant
partons in p+p collisions.

The $p_T$ distribution contributed by two participant partons is
the convolution of two TP-like functions. That is~\cite{43}
\begin{align}
f(p_T) &=\int_0^{p_T}{f_1(p_{t1})}f_2(p_T-p_{t1})dp_{t1} \nonumber\\
&=\int_0^{p_T}{f_2(p_{t2})}f_1(p_T-p_{t2})dp_{t2}.
\end{align}
If Eq. (3) fits the $p_T$ spectrum at the particle level in which
the rest mass of particle is used, Eq. (5) fits the $p_T$ spectrum
at the quark or parton level in which the constituent mass of
quark is used. Although Eq. (3) can fit the $p_T$ spectrum at the
particle level satisfactorily, Eq. (5) can fit the $p_T$ spectrum
at the quark level more significantly. Since Eq. (5) describes the
contributions of participant quarks or partons, we may call it the
result of the participant quark model. In fact, Eq. (5), i.e. the
convolution, is also a result in the framework of the multi-source
thermal model~\cite{44,45}.

\subsection{The revised Tsallis-like function and the root-sum-of-squares of two functions}

According to refs.~\cite{6,26,27,48,48a,48b}, we know that, at
mid-rapidity, the Tsallis-like distribution obeyed by $p_T$ can be
written as
\begin{align}
f(p_T) =& \,\, Cp_Tm_T \nonumber\\
&\times\left[1+\frac{{(q-1)(m_T-\mu-m_0)}}{T}\right]^{-q/(q-1)},
\end{align}
where $\mu$ is the chemical potential, $q$ is the entropy index
that describes the degree of equilibrium, and $C$ is the
normalization constant which is different from those in Eqs. (2)
and (3). Equation (6) is very similar to Eq. (2), if not equal.
Comparing with Eq. (2), Eq. (6) is more in line with the
requirement of thermodynamic consistency due to the fact that
$q/(q-1)$ in Eq. (6) replaced $n_0=1/(q-1)$ in Eq. (2), regardless
of the value of $\mu$. The meanings of the indexes $q$ and $n_0$
are consistent. The closer the index $q$ to 1 is (the larger the
index $n_0$ is), the larger the degree of equilibrium of the
system is.

Similar to Eq. (2), in fitting the $p_T$ spectrum, Eq. (6) is not
ideal in very-low-$p_T$ region, too. Empirically, we may revise
Eq. (6) to the following form:
\begin{align}
f(p_T) =& \,\, C(p_Tm_T)^{a_0} \nonumber\\
&\times\left[1+\frac{{(q-1)(m_T-\mu-m_0)}}{T}\right]^{-q/(q-1)}.
\end{align}
It is similar to Eq. (3). Equation (7) is called the revised
Tsallis-like function by us, to distinguish it from the TP-like
function. For the $i$-th parton, we have
\begin{align}
f_i(p_{ti}) =& \,\, C(p_{ti}m_{ti})^{a_0} \nonumber\\
&\times\left[1+\frac{{(q-1)(m_{ti}-\mu_i-m_{0i})}}{T}\right]^{-q/(q-1)},
\end{align}
where $\mu_i$ denotes the chemical potential of the $i$-th parton.

Of course we can do the convolution of two Eqs. (8). The result is
similar to the convolution of two Eqs. (4), regardless of the
value of $\mu_i$. Although $\mu_i$ causes $m_{0i}$ to increase and
different parameters can be obtained, we may combine $\mu_i$ with
$m_{0i}$ and use smaller $m_{0i}$ to reduce the influence of
$\mu_i$. For other possibilities on the synthesis of $p_{t1}$ and
$p_{t2}$, we give up to use the convolution of two Eqs. (8).
Instead, we consider the root-sum-of-squares of two Eqs. (8) in
which the vectors ${\bm p_{t1}}$ and ${\bm p_{t2}}$ are
perpendicular, and they are the two components of the vector ${\bm
p_{T}}$. That is $p_T=\sqrt{p_{t1}^2+p_{t2}^2}$. This relation is
different from $p_T=p_{t1}+p_{t2}$ discussed in the previous
subsection in which the vectors ${\bm p_{t1}}$ and ${\bm p_{t2}}$
are parallel.

Let $\phi$ denote the azimuth angle of the vector ${\bm p_{T}}$
relative to the vector ${\bm p_{t1}}$. According to
refs.~\cite{43,49}, $p_{t1}=p_T\cos\phi$, $p_{t2}=p_T\sin\phi$.
Therefore, we can obtain the unified probability density function
of $p_T$ and $\phi$ as
\begin{align}
f_{p_T,\phi}(p_T,\phi) &= p_Tf_{1,2}(p_{t1},p_{t2}) \nonumber\\
&= p_T f_1(p_{t1})f_2(p_{t2}) \nonumber\\
&= p_T f_1(p_T\cos\phi)f_2(p_T\sin\phi).
\end{align}
The probability density function of $p_T$ is
\begin{align}
f(p_T)&= \int_0^{2\pi}f_{p_T,\phi}(p_T,\phi)d\phi \nonumber\\
&=p_T\int_0^{2\pi}f_1(p_T\cos\phi)f_2(p_T\sin\phi)d\phi.
\end{align}

We now discuss the chemical potential $\mu_i$ in Eq. (8) that is
used in Eq. (10). It is known that the chemical potential has a
great influence on $p_T$ spectrum at low energy, but has a little
influence at high energy~\cite{51,52,53,54,55,56,57}. According to
refs.~\cite{58,59,60}, the chemical potential $\mu_B$ of baryons
in central nucleus-nucleus collisions is related to the
center-of-mass energy $\sqrt{s_{NN}}$ per nucleon pair.
Empirically, one has
\begin{align}
\mu_B=\frac{1.303\,(\rm GeV)}{1+0.286\sqrt{s_{NN}}/{\rm GeV}}.
\end{align}
It is known from ref.~\cite{51} that $\mu_i=\mu_B/3$ because a
baryon consists of three quarks. In p+p collisions, we have
approximately the same $\mu_B$ as given is Eq. (11). In this
paper, we consider two cases, $\mu_i=\mu_B/3$ and $\mu_i=0$, for
checking the influence of non-zero chemical potential on $p_T$
spectrum.

In the above discussions, the case of mid-rapidity is used. This
means that we have used $y\approx0$ for a narrow rapidity range
which covers the mid-rapidity. If the narrow rapidity range does
not cover the mid-rapidity, one may simply shift it to cover the
mid-rapidity by adding or deducting a value, based on the
additivity of rapidity. This treatment is used to exclude the
kinetic energy of directed motion of the source, which should not
contribute to the temperature. If the rapidity range is not too
narrow, e.g. it is larger than three rapidity units, one may
consider an integral for $y$. Simply, one may replace the
transverse mass in the above equations by the energy obtained by
the product of the transverse mass and $\cosh y$. For the narrow
rapidity range, considering the integral for $y$ may cause mainly
the variation of normalization and does not cause the free
parameters to change significantly.

In this work, the main modification for the TP-like function and
the Tsallis-like distribution is the introduction of correction
index $a_0$. This introduction seems to cause the variation of
thermodynamic consistency~\cite{60a}. However, we may think that
the thermodynamic consistency is mainly determined by the
remainder excluded the added item $(p_Tm_T)^{a_0-1}$ or
$(p_{ti}m_{ti})^{a_0-1}$ in Eqs. (3), (4), (7), and (8). The
influence of $(p_Tm_T)^{a_0-1}$ or $(p_{ti}m_{ti})^{a_0-1}$ can be
combined into the normalization. This combination also includes
the dimension. As a result, the dimensions for the two sides of
the equations are finally the same. As mentioned in the last
subsection, both the units in the two sides of the equations are
$({\rm GeV}/c)^{-1}$, if the unit of $p_T$ is ${\rm GeV}/c$.

In addition, as free parameters, the values of $q$ and $a_0$ are
determined by the $p_T$ spectra. If the $p_T$ spectra can be
approximately fitted by the Boltzmann--Gibbs distribution, we have
naturally $q\rightarrow1$ and $a_0\rightarrow1$. Then, the
Boltzmann--Gibbs distribution is recovered. If the $p_T$ spectra
cannot be fitted by the Boltzmann--Gibbs distribution, we have
$q\neq1$ and any $a_0$, or $q\rightarrow1$ and $a_0\neq1$. It does
not matter whether the Boltzmann--Gibbs distribution is recovered
or not. It depends on the values of parameters themselves. In the
particular case where $q\rightarrow1$ and $a_0\neq1$, the
Boltzmann--Gibbs distribution is modified. In the particular case
where $q\neq1$ and $a_0\rightarrow1$, the TP-like function and the
Tsallis-like distribution are recovered.

\subsection{An energy independent scaling --- the $x_T$ scaling}

We know that the prompt photons refer to all photons directly
generated by hard processes or bremsstrahlung. From the $p_T$
spectra of prompt photons in p+p collisions, it can be seen that
the fragmentation of hard scattering partons is the main mechanism
of high-$p_T$ prompt photons. It is predicted that this production
mechanism may produce a behavior called $x_T$ scaling, where $x_T$
is defined as $x_T=2p_T/\sqrt{s}$. This scaling behavior was
firstly discovered in the preliminary data from the Intersecting
Storage Rings (ISR) of CERN~\cite{79}. Up to now, this scaling
behavior has been used to analyze the transformation and other
properties of the soft and hard processes which produce
particles~\cite{80}.

According to ref.~\cite{80}, in the corresponding $x_T$ range and
for the case of mid-rapidity, the invariant cross-section can be
expressed as
\begin{align}
E\frac{d^3\sigma}{dp^3}=\frac{1}{p_T^{n}}F(x_T)=\frac{1}{\sqrt{s}^n}G(x_T),
\end{align}
where $\sigma$ is the cross-section and $F(x)$ and $G(x)$ are
general scaling functions which will be determined from the quoted
data. Focus in the index $n$, especially noting that this index is
not a constant. In addition, it is different from the power index
in the TP-like function introduced in the beginning of this
section. The value of $n$ depends on the energy $\sqrt{s_i}$ of
the constituent parton and the range of $x_T$. At least, $n$ is
related to $\sqrt{s}$.

To know the value of $n$, one may use two methods based on Eq.
(12)~\cite{80}: (i) Using the ratio of logarithmic yield ratio to
logarithmic energy ratio
\begin{align}
n=-\frac{\log[{\rm yield}(x_T,\sqrt{s_1})/{\rm
yield}(x_T,\sqrt{s_2})]}{\log(\sqrt{s_1}/\sqrt{s_2})};
\end{align}
(ii) Fitting the $x_T$ distribution in terms of invariant
cross-section at different energies,
\begin{align}
E\frac{d^3\sigma}{dp^3}=\left(\frac{B}{\sqrt{s}/{\rm
GeV}}\right)^{n} x_T^{-m},
\end{align}
where $B$ and $m$ are parameters related to $\sqrt{s}$ and $x_T$.
In a real treatment, the form of
\begin{align}
(\sqrt{s}/{\rm GeV})^nE\frac{d^3\sigma}{dp^3}=G(x_T)
\end{align}
is used to fit the spectra of $x_T$ scaling, where the general
scaling function $G(x_T)$ is pending to be determined.

\section{Results and discussion}

\subsection{Comparison with data}

In the above section, we have used the probability density
function, $f(p_T)=(1/N)dN/dp_T$. The experimental data quoted in
this paper are in the form of invariant cross-section,
$Ed^3\sigma/dp^3=(1/2\pi p_T)d^2\sigma/dp_Tdy$. Thus, one has the
relation $(1/2\pi p_T)\sigma_0 f(p_T)/dy=Ed^3\sigma/dp^3$. Here,
$f(p_T)/dy$ only means that $f(p_T)$ is divided by a concrete
amount $dy$, and $\sigma_0$ is the normalization constant which is
used in the calculations to compare with the data. If the
experimental data are in other forms, one may adjust the relation
conveniently. For examples, one has at least the following
relations: $(1/2\pi p_T) N_0f(p_T)/dy=Ed^3N/dp^3$, $\sigma_0
f(p_T)/dy=d^2\sigma/dp_Tdy$, $N_0f(p_T)/dy=d^2N/dp_Tdy$, $\sigma_0
f(p_T)=d\sigma/dp_T$, and $N_0f(p_T)=dN/dp_T$, where the
representations in the right side of these equations are various
possible forms used in experiments, and $N_0$ is the normalization
constant used to compare with the data.

As mentioned above, the function form of $f(p_T)=(1/N)dN/dp_T$ is
used in the calculations. This renders that $f(p_T)$ is normalized
to 1. The normalization constant in $f(p_T)$ is not related to
$\sigma_0$ and $N_0$. The latter two are related to the data in
the comparison. In addition, in the distributions based on
non-extensive statistics satisfying thermodynamic consistency
relations, the normalization which is proportional to the volume
$V$ and the degenerecency factor $g$ is re-normalized in $f(p_T)$
because the probability density function is needed for us. After
the re-normalization, although the volume no longer appears, its
meaning can be also reflected by $N_0$. The larger $N_0$ is, the
larger the volume is.

Figure 1(a) shows the invariant cross-section of direct photons
produced in two-body processes in p+p collisions at different
collision energies with different mid-$\eta$ ranges. The points
(symbols) in different colors in the picture represent different
rapidities, energies, and collaborations. For clarity, the results
(symbols) corresponding to the UA6 Collaboration at 24.3 GeV with
rapidity $-0.1<|\eta|<0.9$~\cite{61}, the R806 Collaboration at 63
GeV with rapidity $|\eta|<0.2$~\cite{62}, and the R110
Collaboration at 63 GeV with rapidity $|\eta|<0.8$~\cite{63} are
multiplied by the factors $10^{3}$, $10^{-2}$, and $10^{-4}$,
respectively. The result corresponding to the CCOR Collaboration
at 62.4 GeV with rapidity $|\eta|<0.45$~\cite{64} is not re-scaled
(i.e. the re-scaling factor is $10^0$ in this case). The solid,
dashed, dot-dashed, and dotted curves in Figure 1(a) represent the
results obtained by Eqs. (3), (5), (10) with $\mu_i=\mu_B/3$, and
(10) with $\mu_i=0$, respectively. Because Eqs. (3) and (7) are
similar to each other, only the result from Eq. (3) is presented.

\begin{figure*}[htbp]
\begin{center}
\includegraphics{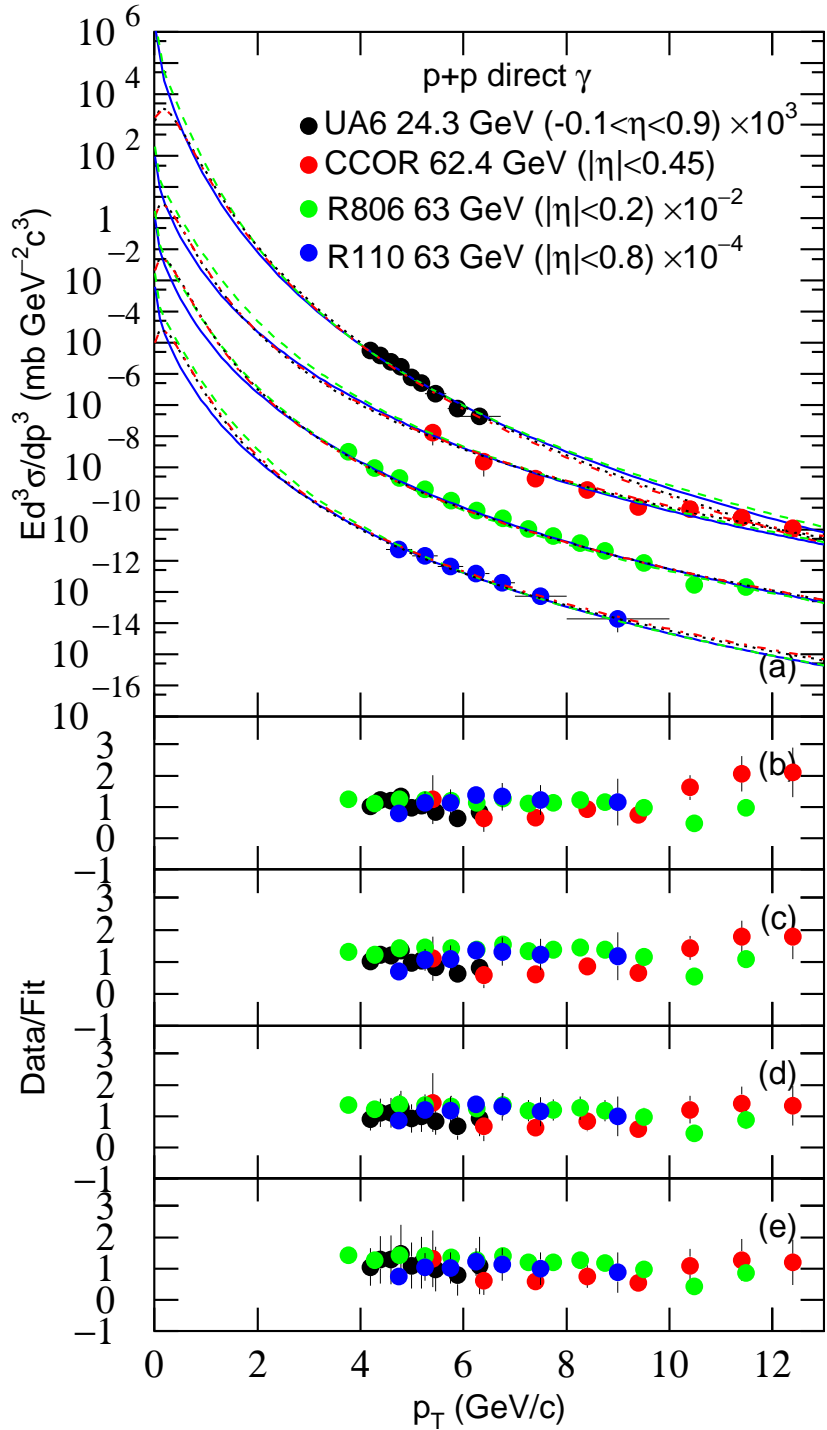}
\end{center}
\caption{\small (a) The invariant cross-section of direct photons
produced in two-body processes in p+p collisions at 24.3, 62.4,
and 63 GeV. The symbols with different colors represent the
results from different rapidity intervals and measured by the UA6,
CCOR, R806, R110 Collaborations~\cite{61,62,63,64}. For clarity,
the symbols are re-scaled by certain factors. The solid, dashed,
dot-dashed, and dotted curves represent the results obtained by
Eqs. (3), (5), (10) with $\mu_i=\mu_B/3$, and (10) with $\mu_i=0$,
respectively. (b)--(e) The ratios of data to fit obtained from
Eqs. (3), (5), (10) with $\mu_i=\mu _B/3$, and (10) with
$\mu_i=0$, respectively.}
\end{figure*}

\begin{figure*}[htbp]
\begin{center}
\includegraphics{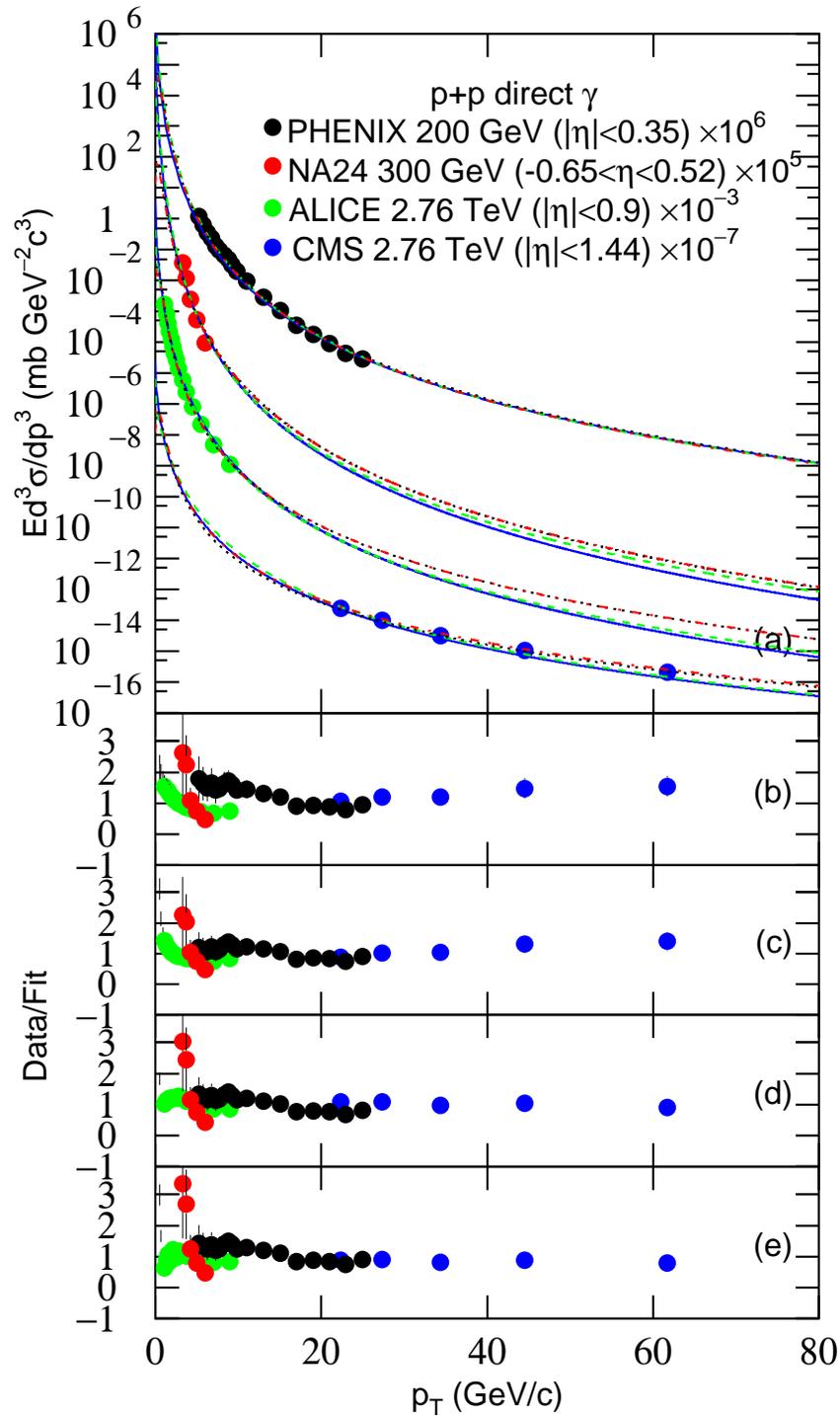}
\end{center}
\caption{\small Same as in Figure 1, but showing the results at
200 GeV, 300 GeV, and 2.76 TeV. The symbols in Figure 2(a)
represent the data measured by the PHENIX, NA24, CMS, and ALICE
Collaborations~\cite{65,66,67,68}.}
\end{figure*}

\begin{figure*}[htbp]
\begin{center}
\includegraphics{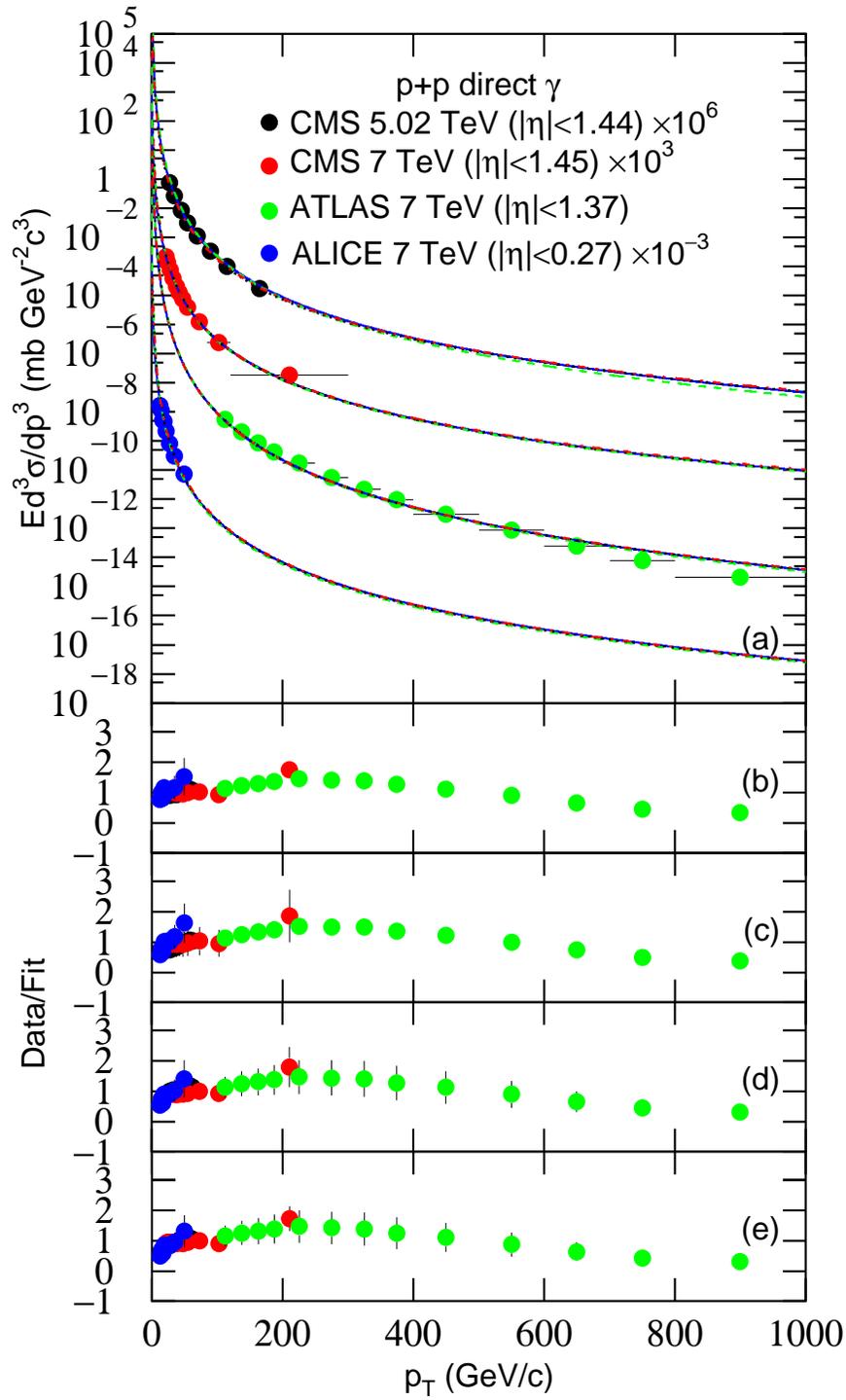}
\end{center}
\caption{\small Same as in Figure 1, but showing the results at
5.02 and 7 TeV. The symbols in Figure 3(a) represent the data
measured by the CMS, ATLAS, and ALICE
Collaborations~\cite{69,70,71,73}.}
\end{figure*}

\begin{figure*}[htbp]
\begin{center}
\includegraphics{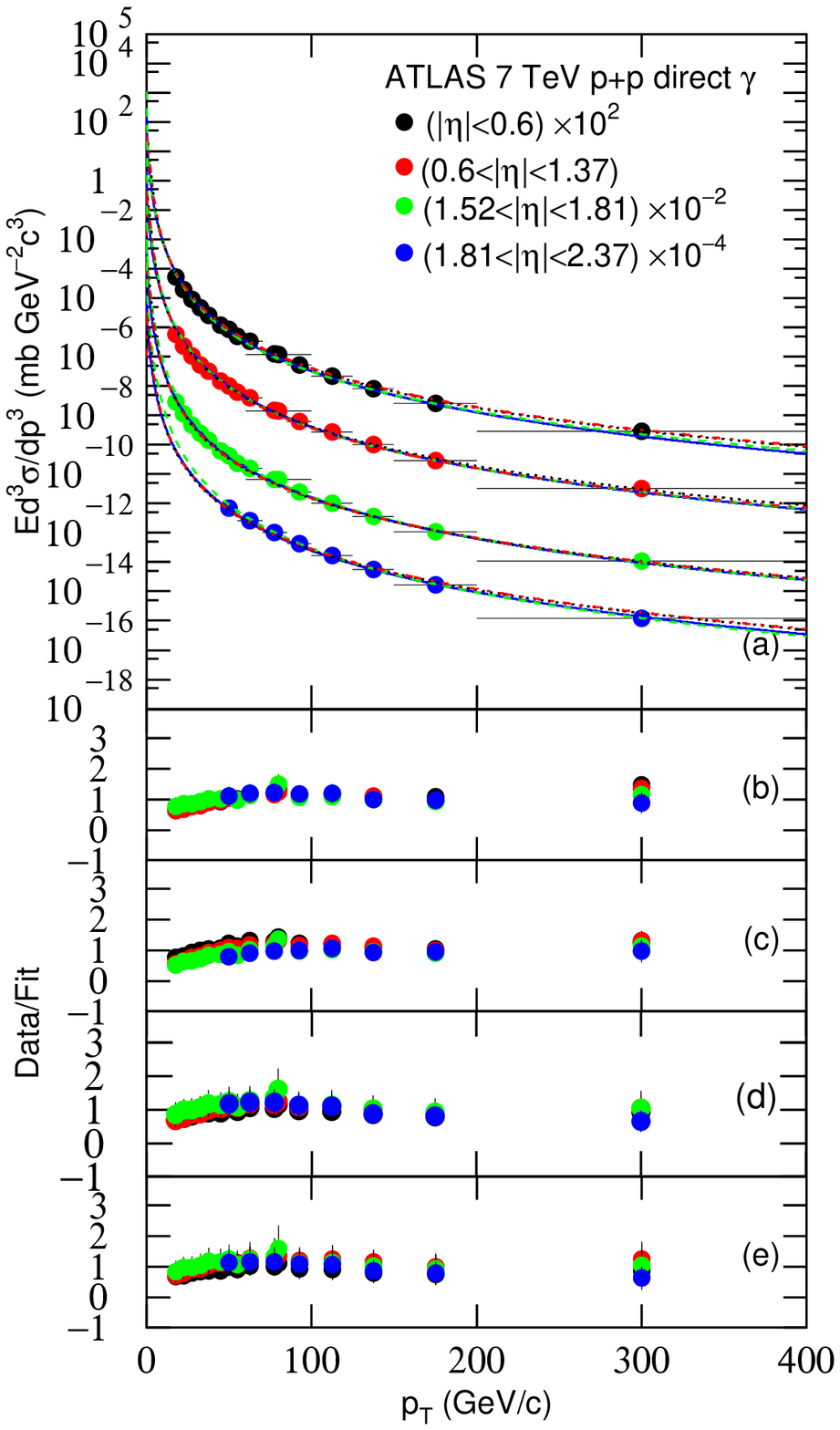}
\end{center}
\caption{\small Same as in Figure 1, but showing the results at 7
TeV. The symbols in Figure 4(a) represent the data measured by the
ATLAS Collaboration~\cite{72,74}.}
\end{figure*}

\begin{figure*}[htbp]
\begin{center}
\includegraphics{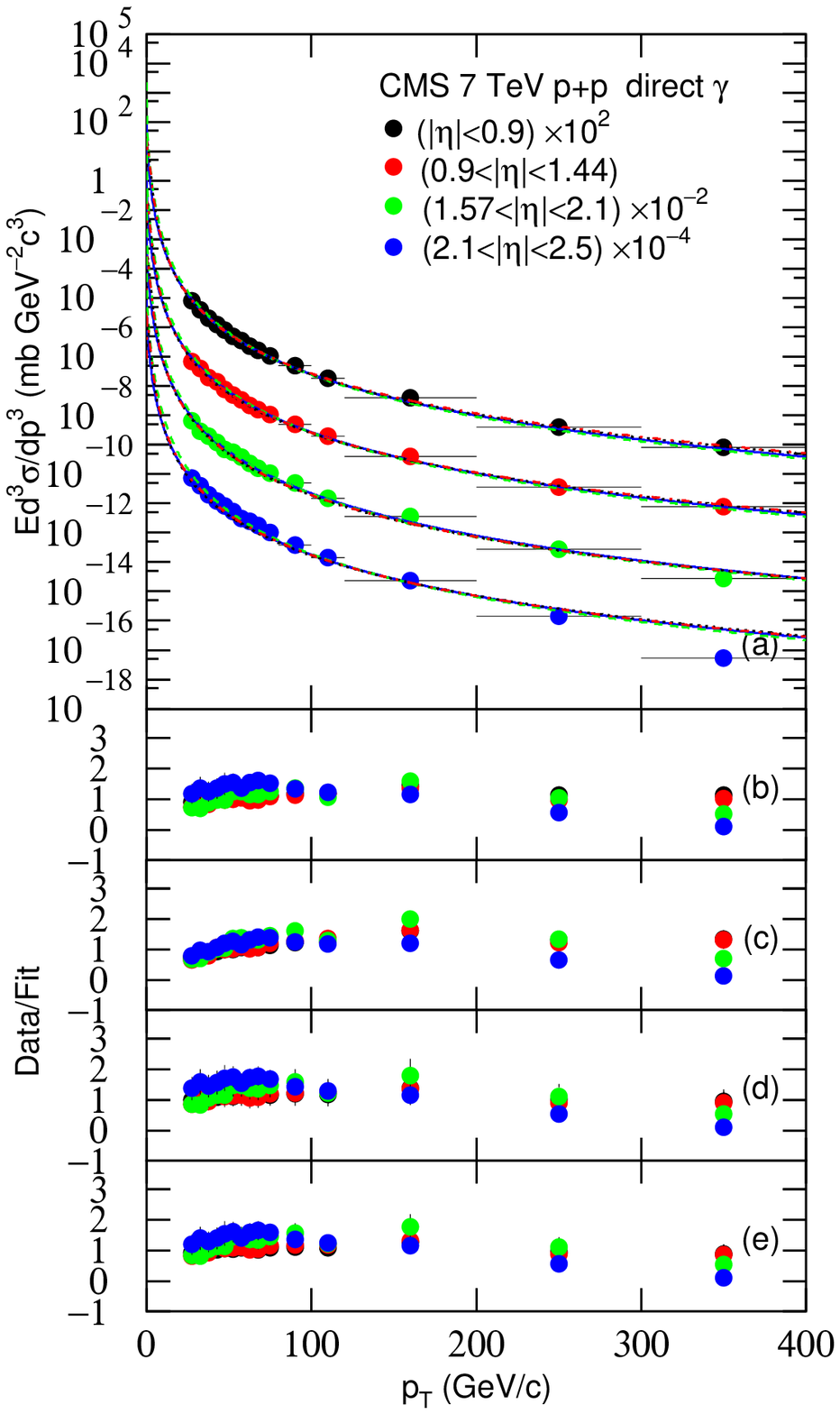}
\end{center}
\caption{\small Same as in Figure 1, but showing the results at 7
TeV. The symbols in Figure 5(a) represent the data measured by the
CMS Collaboration~\cite{75}.}
\end{figure*}

\begin{figure*}[htbp]
\begin{center}
\includegraphics{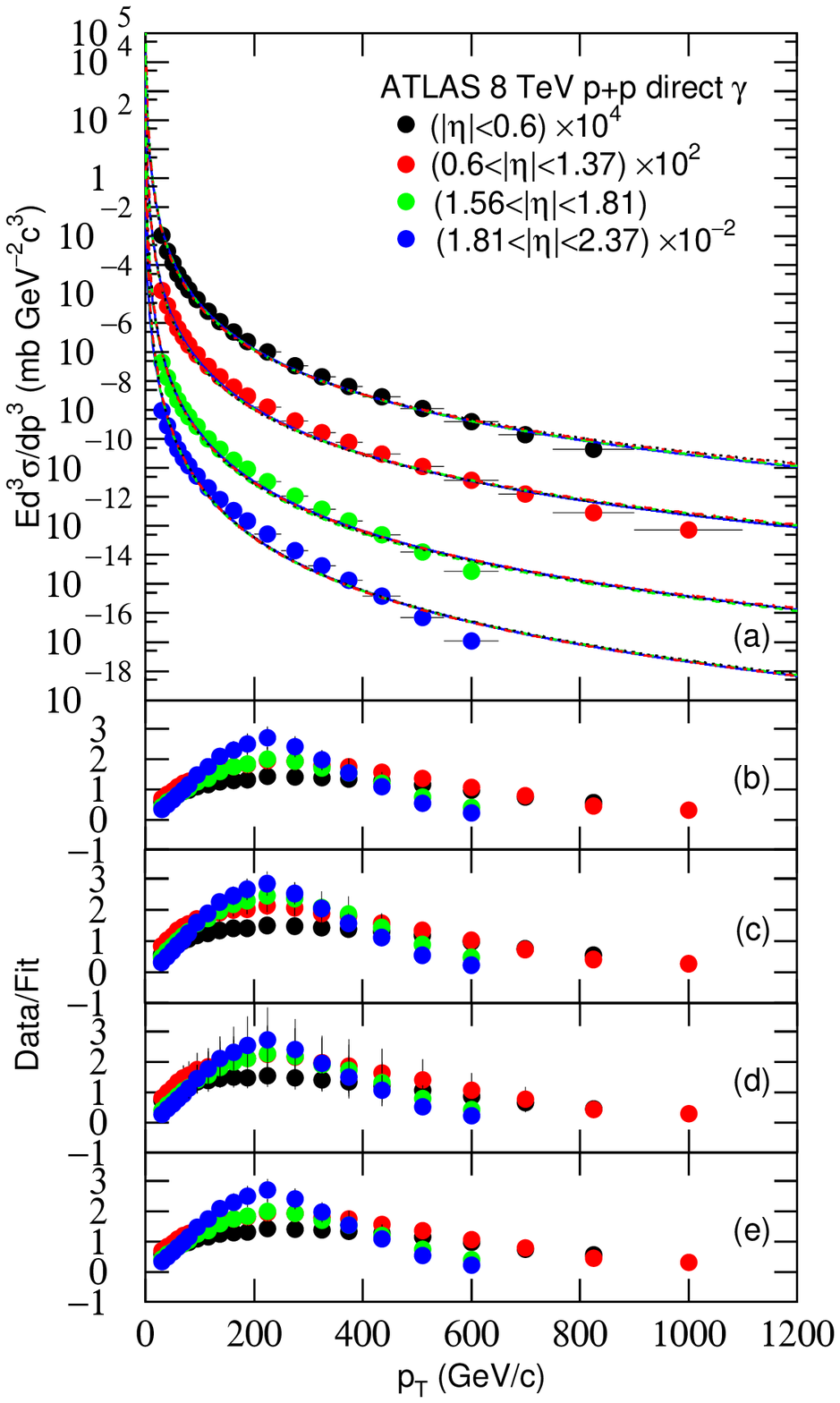}
\end{center}
\caption{\small Same as in Figure 1, but showing the results at 8
TeV. The symbols in Figure 6(a) represent the data measured by the
ATLAS Collaboration~\cite{76}.}
\end{figure*}

\begin{figure*}[htbp]
\begin{center}
\includegraphics{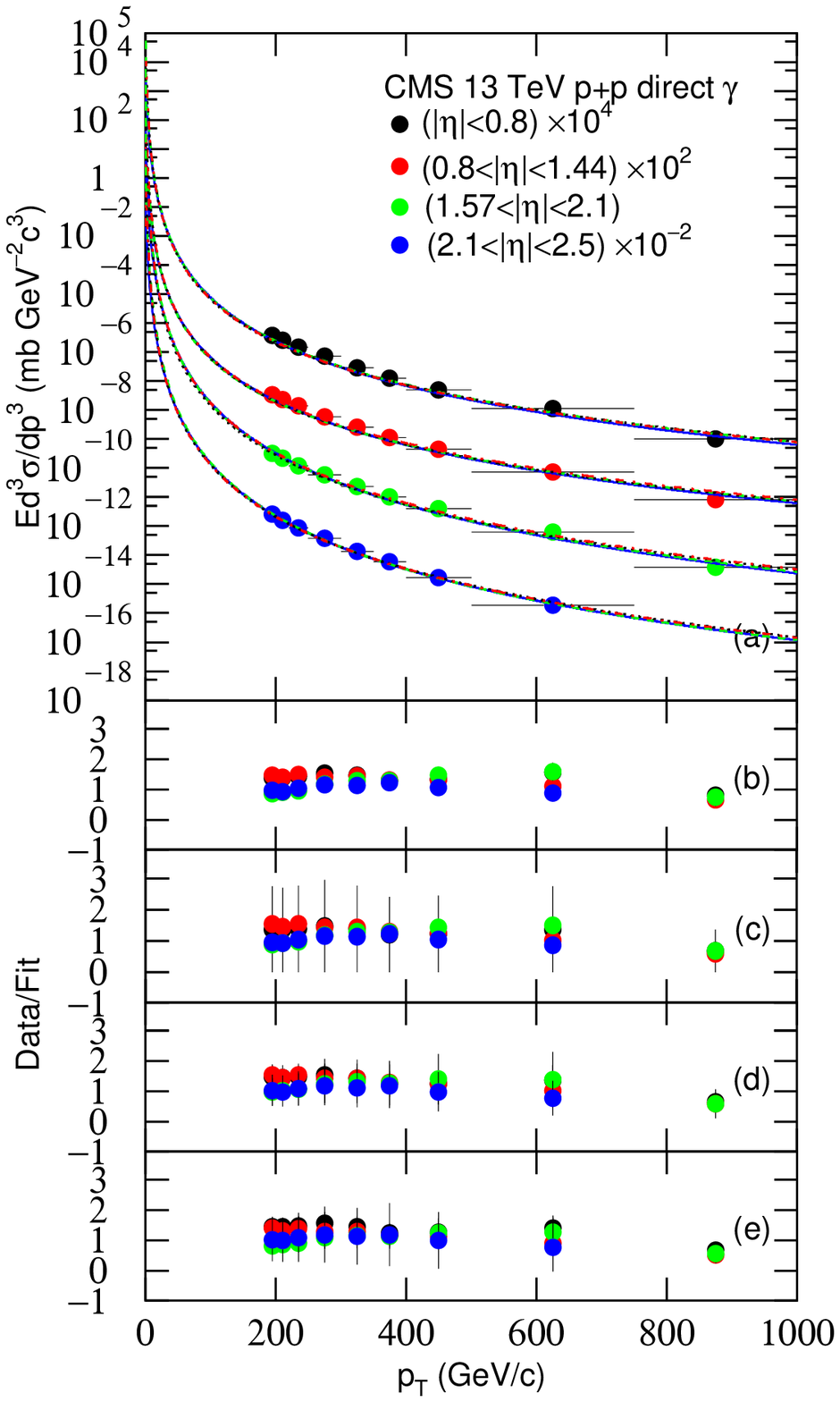}
\end{center}
\caption{\small Same as in Figure 1, but showing the results at 13
TeV. The symbols in Figure 7(a) represent the data measured by the
CMS Collaboration~\cite{77}.}
\end{figure*}

\begin{figure*}[htbp]
\begin{center}
\includegraphics{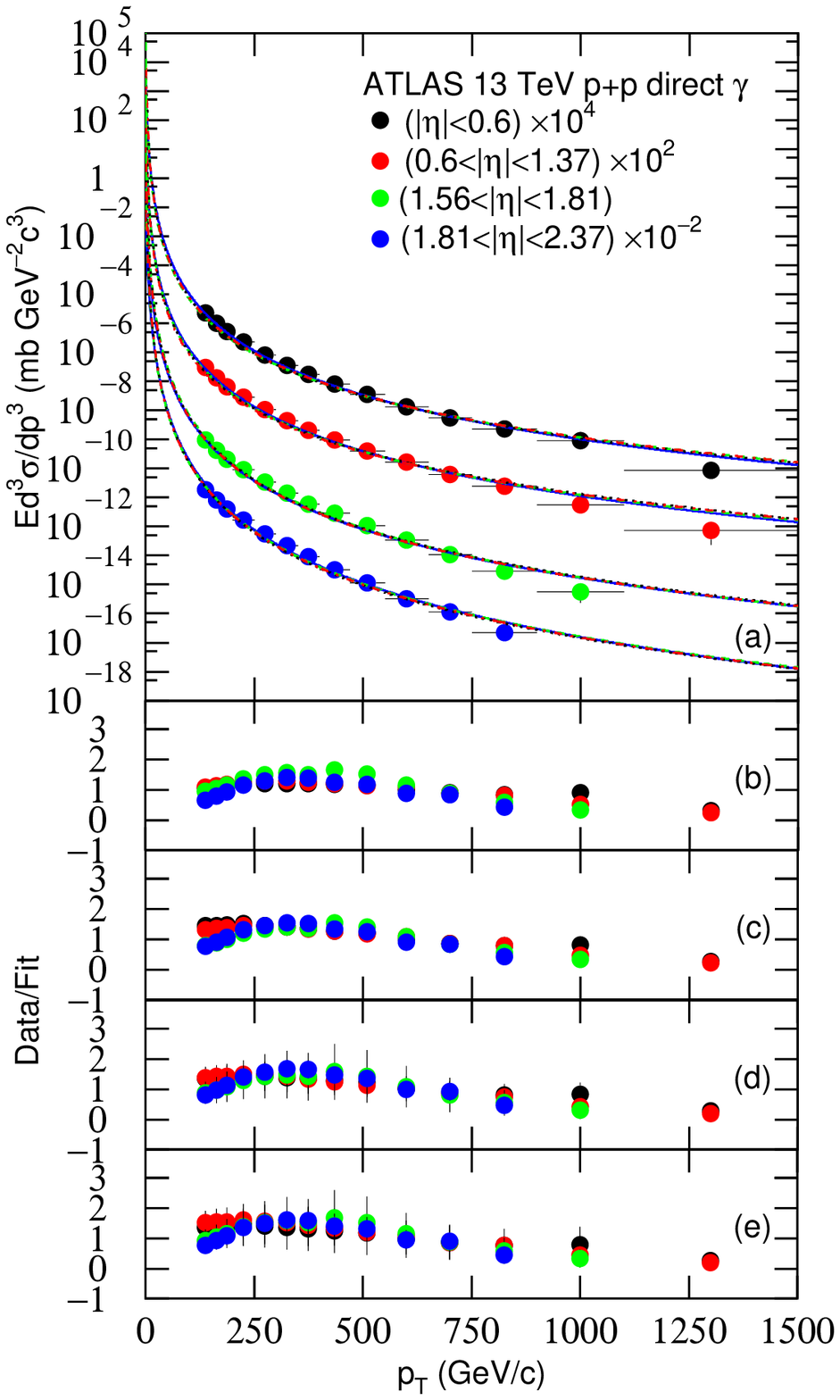}
\end{center}
\caption{\small Same as in Figure 1, but showing the results at 13
TeV. The symbols in Figure 8(a) represent the data measured by the
ATLAS Collaboration~\cite{78}.}
\end{figure*}

\begin{table*}[htbp]
\vspace{.0cm} \justifying\noindent {\small Table 1. Values of $T$,
$n_0$, $a_0$, $\sigma_0$, $\chi^2$, and ndof corresponding to the
solid curves in Figures 1(a)--8(a), which are fitted by the
TP-like function [Eq. (3)].} \vspace{0mm}
\begin{center}
{\small
\begin{tabular}{lllllllll}
\hline \hline
Figure & Collab. & Energy & $\eta$ & $T$ (GeV) & $n_0$ & $a_0$ & $\sigma_0$ (mb) & $\chi^2$/ndof\\
\hline
Figure 1(a) & UA6  & 24.3 GeV & $-0.1<\eta<0.9$   & $0.088\pm0.001$ & $12.921\pm0.004$& $0.301\pm0.003$ & $0.365\pm0.004$  & $8/5$ \\
            & CCOR & 62.4 GeV & $|\eta|<0.45$     & $0.125\pm0.001$ & $9.771\pm0.018$ & $0.330\pm0.003$ & $0.210\pm0.003$  & $13/4$ \\
            & R806 & 63 GeV   & $|\eta|<0.2$      & $0.124\pm0.002$ & $9.532\pm0.016$ & $0.331\pm0.002$ & $0.197\pm0.005$  & $119/10$ \\
            & R110 & 63 GeV   & $|\eta|<0.8$      & $0.124\pm0.001$ & $9.291\pm0.032$ & $0.331\pm0.003$ & $0.498\pm0.005$  & $5/3$ \\
\hline
Figure 2(a) & PHENIX& 200 GeV & $|\eta|<0.35$     & $0.148\pm0.001$ & $6.223\pm0.003$ & $0.354\pm0.002$ & $0.102\pm0.002$  & $33/14$ \\
            & NA24  & 300 GeV & $-0.65<\eta<0.52$ & $0.098\pm0.003$ & $7.105\pm0.003$ & $0.146\pm0.002$ & $0.051\pm0.002$  & $11/1$ \\
            & CMS   & 2.76 TeV& $|\eta|<1.44$     & $0.258\pm0.003$ & $4.583\pm0.084$ & $0.378\pm0.002$ & $4.142\pm0.011$  & $7/1$ \\
            & ALICE & 2.76 TeV& $|\eta|<0.9$      & $0.165\pm0.001$ & $6.200\pm0.003$ & $0.224\pm0.002$ & $7.701\pm0.116$  & $122/14$\\
\hline
Figure 3(a) & CMS   & 5.02 TeV& $|\eta|<1.44$     & $0.258\pm0.001$ & $4.071\pm0.003$ & $0.372\pm0.002$ & $0.827\pm0.005$  & $8/4$ \\
            & CMS   & 7 TeV   & $|\eta|<1.45$     & $0.546\pm0.001$ & $4.125\pm0.003$ & $0.596\pm0.002$ & $0.883\pm0.002$  & $14/7$ \\
            & ATLAS & 7 TeV   & $|\eta|<1.37$     & $0.565\pm0.001$ & $4.823\pm0.004$ & $0.390\pm0.003$ & $0.746\pm0.008$  & $98/9$ \\
            & ALICE & 7 TeV   & $|\eta|<0.27$     & $0.489\pm0.003$ & $4.418\pm0.002$ & $0.560\pm0.002$ & $0.316\pm0.004$  & $5/5$ \\
\hline
Figure 4(a) & ATLAS & 7 TeV   & $|\eta|<0.6$      & $0.549\pm0.003$ & $4.323\pm0.001$ & $0.600\pm0.002$ & $0.640\pm0.005$  & $17/12$ \\
            &       &         & $0.6<|\eta|<1.37$ & $0.549\pm0.001$ & $4.323\pm0.003$ & $0.600\pm0.002$ & $0.760\pm0.004$  & $12/12$ \\
            &       &         & $1.52<|\eta|<1.81$& $0.549\pm0.001$ & $4.323\pm0.008$ & $0.600\pm0.003$ & $0.261\pm0.003$  & $6/12$ \\
            &       &         & $1.81<|\eta|<2.37$& $0.550\pm0.002$ & $4.397\pm0.001$ & $0.601\pm0.002$ & $0.945\pm0.004$  & $2/4$ \\
\hline
Figure 5(a) & CMS   & 7 TeV   & $|\eta|<0.9$      & $0.549\pm0.001$ & $4.323\pm0.026$ & $0.600\pm0.002$ & $0.560\pm0.012$  & $12/11$ \\
            &       &         & $0.9<|\eta|<1.44$ & $0.559\pm0.001$ & $4.323\pm0.025$ & $0.600\pm0.003$ & $0.619\pm0.005$  & $10/11$ \\
            &       &         & $1.57<|\eta|<2.1$ & $0.549\pm0.001$ & $4.463\pm0.001$ & $0.600\pm0.003$ & $0.548\pm0.006$  & $30/11$ \\
            &       &         & $2.1<|\eta|<2.5$  & $0.549\pm0.001$ & $4.343\pm0.004$ & $0.600\pm0.002$ & $0.100\pm0.003$  & $126/11$ \\
\hline
Figure 6(a) & ATLAS & 8 TeV   & $|\eta|<0.6$      & $0.681\pm0.001$ & $4.931\pm0.003$ & $0.674\pm0.002$ & $2.705\pm0.013$  & $90/16$ \\
            &       &         & $0.6<|\eta|<1.37$ & $0.611\pm0.002$ & $4.902\pm0.005$ & $0.584\pm0.003$ & $6.565\pm0.003$  & $189/16$ \\
            &       &         & $1.36<|\eta|<1.81$& $0.626\pm0.001$ & $5.374\pm0.001$ & $0.671\pm0.003$ & $1.526\pm0.003$  & $109/14$ \\
            &       &         & $1.81<|\eta|<2.37$& $0.686\pm0.001$ & $5.874\pm0.014$ & $0.674\pm0.002$ & $7.912\pm0.012$  & $480/14$ \\
\hline
Figure 7(a) & CMS   & 13 TeV  & $|\eta|<0.8$      & $0.733\pm0.001$ & $4.993\pm0.005$ & $0.854\pm0.002$ & $2.640\pm0.013$  & $46/5$ \\
            &       &         & $0.8<|\eta|<1.44$ & $0.713\pm0.001$ & $4.934\pm0.001$ & $0.854\pm0.003$ & $1.741\pm0.007$  & $43/5$ \\
            &       &         & $1.57<|\eta|<2.1$ & $0.703\pm0.001$ & $5.837\pm0.005$ & $0.864\pm0.002$ & $3.520\pm0.015$  & $17/5$ \\
            &       &         & $2.1<|\eta|<2.5$  & $0.703\pm0.001$ & $6.033\pm0.005$ & $0.844\pm0.001$ & $4.880\pm0.012$  & $3/4$ \\
\hline
Figure 8(a) & ATLAS & 13 TeV  & $|\eta|<0.6$      & $0.731\pm0.001$ & $4.933\pm0.005$ & $0.854\pm0.002$ & $2.441\pm0.013$  & $51/10$ \\
            &       &         & $0.6<|\eta|<1.37$ & $0.793\pm0.001$ & $4.953\pm0.018$ & $0.812\pm0.003$ & $3.833\pm0.011$  & $61/10$ \\
            &       &         & $1.56<|\eta|<1.81$& $0.793\pm0.020$ & $5.418\pm0.008$ & $0.815\pm0.002$ & $2.533\pm0.014$  & $47/9$ \\
            &       &         & $1.81<|\eta|<2.37$& $0.793\pm0.002$ & $5.804\pm0.001$ & $0.615\pm0.013$ & $5.622\pm0.012$  & $52/8$ \\
\hline
\end{tabular}}
\end{center}
\end{table*}

\begin{table*}[htbp]
\vspace{.0cm} \justifying\noindent {\small Table 2. Values of $T$,
$n_0$, $a_0$, $\sigma_0$, $\chi^2$, and ndof corresponding to the
dashed curves in Figures 1(a)--8(a), which are fitted by the
convolution of two TP-like functions [Eq. (5)].} \vspace{0mm}
\begin{center}
{\small
\begin{tabular}{lllllllll}
\hline \hline
Figure & Collab. & Energy & $\eta$ & $T$ (GeV) & $n_0$ & $a_0$ & $ \sigma_0$ (mb) & $\chi^2$/ndof\\
\hline
Figure 1(a) & UA6   & 24.3 GeV& $-0.1<\eta<0.9$   & $0.097\pm0.001$ & $10.720\pm0.014$& $-0.337\pm0.003$ & $0.284\pm0.007$  & 9/5 \\
            & CCOR  & 62.4 GeV& $|\eta|<0.45$     & $0.131\pm0.002$ & $8.120\pm0.017$ & $-0.259\pm0.002$ & $0.244\pm0.005$  & 15/4 \\
            & R806  & 63 GeV  & $|\eta|<0.2$      & $0.130\pm0.001$ & $7.843\pm0.009$ & $-0.265\pm0.003$ & $0.100\pm0.002$  & 125/11 \\
            & R110  & 63 GeV  & $|\eta|<0.8$      & $0.130\pm0.002$ & $7.811\pm0.024$ & $-0.271\pm0.001$ & $0.396\pm0.005$  & 8/3 \\
\hline
Figure 2(a) & PHENIX& 200 GeV & $|\eta|<0.35$     & $0.202\pm0.001$ & $5.543\pm0.004$ & $-0.176\pm0.003$ & $0.073\pm0.002$  & 92/14 \\
            & NA24  & 300 GeV & $-0.65<\eta<0.52$ & $0.129\pm0.001$ & $6.153\pm0.004$ & $-0.195\pm0.002$ & $0.038\pm0.002$  & 10/1 \\
            & CMS   & 2.76 TeV& $|\eta|<1.44$     & $0.469\pm0.002$ & $3.945\pm0.003$ & $-0.139\pm0.002$ & $0.619\pm0.006$  & 3/1 \\
            & ALICE & 2.76 TeV& $|\eta|<0.9$      & $0.239\pm0.001$ & $5.183\pm0.003$ & $-0.229\pm0.002$ & $2.580\pm0.004$  & 118/14 \\
\hline
Figure 3(a) & CMS   & 5.02 TeV& $|\eta|<1.44$     & $0.503\pm0.003$ & $3.580\pm0.006$ & $-0.132\pm0.001$ & $0.771\pm0.005$  & 20/4 \\
            & CMS   & 7 TeV   & $|\eta|<1.45$     & $0.595\pm0.001$ & $3.263\pm0.012$ & $-0.128\pm0.003$ & $0.624\pm0.003$  & 18/7 \\
            & ATLAS & 7 TeV   & $|\eta|<1.37$     & $0.566\pm0.001$ & $4.244\pm0.002$ & $-0.120\pm0.003$ & $0.776\pm0.006$  & 87/9 \\
            & ALICE & 7 TeV   & $|\eta|<0.27$     & $0.359\pm0.002$ & $3.587\pm0.002$ & $-0.124\pm0.001$ & $0.354\pm0.006$  & 23/5 \\
\hline
Figure 4(a) & ATLAS &  7 TeV  & $|\eta|<0.6$      & $0.636\pm0.002$ & $3.231\pm0.003$ & $-0.130\pm0.002$ & $0.662\pm0.005$  & 13/12\\
            &       &         & $0.6<|\eta|<1.37$ & $0.625\pm0.002$ & $3.351\pm0.003$ & $-0.129\pm0.002$ & $0.696\pm0.003$  & 15/12 \\
            &       &         & $1.52<|\eta|<1.81$& $0.619\pm0.001$ & $3.463\pm0.003$ & $-0.132\pm0.001$ & $0.382\pm0.005$  & 22/12 \\
            &       &         & $1.81<|\eta|<2.37$& $0.592\pm0.006$ & $3.731\pm0.012$ & $-0.130\pm0.001$ & $0.910\pm0.023$  & 2/4 \\
\hline
Figure 5(a) & CMS   &  7 TeV  & $|\eta|<0.9$      & $0.596\pm0.005$ & $3.532\pm0.004$ & $-0.120\pm0.002$ & $0.581\pm0.004$  & $27/11$ \\
            &       &         & $0.9<|\eta|<1.44$ & $0.591\pm0.007$ & $3.534\pm0.002$ & $-0.124\pm0.001$ & $0.593\pm0.004$  & 25/11 \\
            &       &         & $1.57<|\eta|<2.1$ & $0.591\pm0.002$ & $3.652\pm0.003$ & $-0.123\pm0.001$ & $0.477\pm0.002$  & 41/11 \\
            &       &         & $2.1<|\eta|<2.5$  & $0.592\pm0.002$ & $3.633\pm0.015$ & $-0.123\pm0.002$ & $0.189\pm0.005$  & 64/11 \\
\hline
Figure 6(a) & ATLAS &  8 TeV  & $|\eta|<0.6$      & $0.602\pm0.002$ & $3.953\pm0.004$ & $-0.106\pm0.002$ & $2.745\pm0.007$  & 106/16\\
            &       &         & $0.6<|\eta|<1.37$ & $0.622\pm0.002$ & $3.943\pm0.003$ & $-0.107\pm0.001$ & $2.740\pm0.019$  & 257/16 \\
            &       &         & $1.36<|\eta|<1.81$& $0.604\pm0.002$ & $4.348\pm0.001$ & $-0.136\pm0.002$ & $2.812\pm0.008$  & 109/14 \\
            &       &         & $1.81<|\eta|<2.37$& $0.604\pm0.003$ & $4.798\pm0.025$ & $-0.128\pm0.002$ & $8.732\pm0.003$  & 527/14 \\
\hline
Figure 7(a) & CMS   &  13 TeV & $|\eta|<0.8$      & $0.772\pm0.001$ & $3.729\pm0.003$ & $-0.109\pm0.002$ & $2.136\pm0.007$  & 39/5 \\
            &       &         & $0.8<|\eta|<1.44$ & $0.731\pm0.001$ & $3.589\pm0.003$ & $-0.101\pm0.001$ & $1.314\pm0.003$  & 48/5 \\
            &       &         & $1.57<|\eta|<2.1$ & $0.734\pm0.002$ & $4.654\pm0.004$ & $-0.098\pm0.001$ & $3.573\pm0.003$  & 15/5 \\
            &       &         & $2.1<|\eta|<2.5$  & $0.731\pm0.002$ & $4.965\pm0.004$ & $-0.098\pm0.001$ & $5.200\pm0.112$  & 3/4 \\
\hline
Figure 8(a) & ATLAS &  13 TeV & $|\eta|<0.6$      & $0.739\pm0.005$ & $3.594\pm0.005$ & $-0.109\pm0.001$ & $0.921\pm0.004$  & 119/10 \\
            &       &         & $0.6<|\eta|<1.37$ & $0.765\pm0.001$ & $3.783\pm0.012$ & $-0.091\pm0.001$ & $2.034\pm0.002$  & 98/10 \\
            &       &         & $1.56<|\eta|<1.81$& $0.784\pm0.001$ & $4.439\pm0.005$ & $-0.101\pm0.001$ & $4.168\pm0.006$  & 40/9 \\
            &       &         & $1.81<|\eta|<2.37$& $0.801\pm0.001$ & $4.845\pm0.003$ & $-0.104\pm0.002$ & $5.310\pm0.123$  & 48/8 \\
\hline
\end{tabular}}
\end{center}
\end{table*}

\begin{table*}[htbp]
\vspace{0.0cm} \justifying\noindent {\small Table 3. Values of
$T$, $q$, $a_0$, $\sigma_0$, $\chi^2$, and ndof corresponding to
the dotted curves in Figures 1(a)--8(a), which are fitted by the
revised Tsallis-like function when $\mu_i=\mu_B/3$ [Eq. (10)].}
\vspace{0mm}
\begin{center}
{\small
\begin{tabular}{lllllllll}
\hline \hline
Figure & Collab. & Energy & $\eta$ & $T$ (GeV) & $q$ & $a_0$ & $\sigma_0$ (mb) & $\chi^2$/ndof\\
\hline
Figure 1(a)& UA6   & 24.3 GeV& $-0.1<\eta<0.9$   & $0.140\pm0.004$ & $1.068\pm0.001$ & $0.229\pm0.002$ & $0.292\pm0.003$ & 5/5 \\
           & CCOR  & 62.4 GeV& $|\eta|<0.45$     & $0.135\pm0.002$ & $1.116\pm0.001$ & $0.230\pm0.001$ & $0.253\pm0.002$ & 12/4 \\
           & R806  & 63 GeV  & $|\eta|<0.2$      & $0.132\pm0.002$ & $1.112\pm0.001$ & $0.231\pm0.002$ & $0.208\pm0.001$ & 163/11 \\
           & R110  & 63 GeV  & $|\eta|<0.8$      & $0.135\pm0.002$ & $1.118\pm0.001$ & $0.230\pm0.001$ & $0.410\pm0.001$ & 4/3 \\
\hline
Figure 2(a)& PHENIX& 200 GeV & $|\eta|<0.35$     & $0.223\pm0.001$ & $1.146\pm0.002$ & $0.309\pm0.002$ & $0.076\pm0.003$ & 26/14 \\
           & NA24  & 300 GeV & $-0.65<\eta<0.52$ & $0.198\pm0.001$ & $1.131\pm0.002$ & $0.308\pm0.001$ & $0.038\pm0.003$ & 14/1 \\
           & CMS   & 2.76 TeV& $|\eta|<1.44$     & $0.234\pm0.002$ & $1.218\pm0.002$ & $0.348\pm0.001$ & $5.641\pm0.086$ & 1/1 \\
           & ALICE & 2.76 TeV& $|\eta|<0.9$      & $0.197\pm0.001$ & $1.182\pm0.003$ & $0.128\pm0.001$ & $9.041\pm0.083$ & 47/14 \\
\hline
Figure 3(a)& CMS   & 5.02 TeV& $|\eta|<1.44$     & $0.344\pm0.004$ & $1.203\pm0.003$ & $0.458\pm0.001$ & $0.584\pm0.004$ & 16/4 \\
           & CMS   & 7 TeV   & $|\eta|<1.45$     & $0.352\pm0.001$ & $1.204\pm0.003$ & $0.467\pm0.002$ & $0.761\pm0.003$ & 22/7 \\
           & ATLAS & 7 TeV   & $|\eta|<1.37$     & $0.350\pm0.002$ & $1.172\pm0.003$ & $0.468\pm0.003$ & $0.721\pm0.011$ & 103/9 \\
           & ALICE & 7 TeV   & $|\eta|<0.27$     & $0.359\pm0.002$ & $1.191\pm0.002$ & $0.457\pm0.002$ & $0.252\pm0.002$ & 34/5 \\
\hline
Figure 4(a)& ATLAS & 7 TeV   & $|\eta|<0.6$      & $0.358\pm0.002$ & $1.207\pm0.002$ & $0.488\pm0.001$ & $0.047\pm0.002$ & 12/12 \\
           &       &         & $0.6<|\eta|<1.37$ & $0.358\pm0.002$ & $1.203\pm0.002$ & $0.488\pm0.002$ & $0.427\pm0.008$ & 6/12 \\
           &       &         & $1.52<|\eta|<1.81$& $0.358\pm0.002$ & $1.201\pm0.003$ & $0.487\pm0.001$ & $0.068\pm0.003$ & 9/12 \\
           &       &         & $1.81<|\eta|<2.37$& $0.358\pm0.001$ & $1.201\pm0.002$ & $0.488\pm0.001$ & $0.235\pm0.008$ & 5/4 \\
\hline
Figure 5(a)& CMS   & 7 TeV   & $|\eta|<0.9$      & $0.358\pm0.003$ & $1.201\pm0.002$ & $0.487\pm0.001$ & $0.350\pm0.006$ & 13/11 \\
           &       &         & $0.9<|\eta|<1.44$ & $0.358\pm0.002$ & $1.201\pm0.003$ & $0.488\pm0.001$ & $0.208\pm0.006$ & 10/11 \\
           &       &         & $1.57<|\eta|<2.1$ & $0.358\pm0.003$ & $1.193\pm0.002$ & $0.488\pm0.002$ & $0.260\pm0.003$ & 47/11 \\
           &       &         & $2.1<|\eta|<2.5$  & $0.358\pm0.001$ & $1.200\pm0.002$ & $0.487\pm0.001$ & $0.112\pm0.003$ & 141/11 \\
\hline
Figure 6(a)& ATLAS & 8 TeV   & $|\eta|<0.6$      & $0.338\pm0.002$ & $1.193\pm0.001$ & $0.388\pm0.001$ & $3.102\pm0.013$ & 127/16 \\
           &       &         & $0.6<|\eta|<1.37$ & $0.353\pm0.002$ & $1.193\pm0.003$ & $0.342\pm0.001$ & $6.420\pm0.006$ & 254/16 \\
           &       &         & $1.56<|\eta|<1.81$& $0.335\pm0.003$ & $1.177\pm0.002$ & $0.348\pm0.001$ & $3.985\pm0.007$ & 120/14 \\
           &       &         & $1.81<|\eta|<2.37$& $0.338\pm0.001$ & $1.163\pm0.002$ & $0.349\pm0.002$ & $6.394\pm0.002$ & 649/14 \\
\hline
Figure 7(a)& CMS   & 13 TeV  & $|\eta|<0.8$      & $0.422\pm0.003$ & $1.192\pm0.002$ & $0.429\pm0.002$ & $2.020\pm0.013$ & 49/5 \\
           &       &         & $0.8<|\eta|<1.44$ & $0.423\pm0.002$ & $1.192\pm0.001$ & $0.423\pm0.003$ & $1.588\pm0.013$ & 48/5 \\
           &       &         & $1.57<|\eta|<2.1$ & $0.421\pm0.003$ & $1.168\pm0.003$ & $0.429\pm0.001$ & $3.640\pm0.012$ & 16/5 \\
           &       &         & $2.1<|\eta|<2.5$  & $0.420\pm0.004$ & $1.160\pm0.001$ & $0.423\pm0.003$ & $4.030\pm0.098$ & 4/4 \\
\hline
Figure 8(a)& ATLAS & 13 TeV  & $|\eta|<0.6$      & $0.423\pm0.004$ & $1.195\pm0.002$ & $0.421\pm0.002$ & $1.610\pm0.004$ & 101/10 \\
           &       &         & $0.6<|\eta|<1.37$ & $0.421\pm0.002$ & $1.200\pm0.002$ & $0.422\pm0.001$ & $2.820\pm0.014$ & 121/10 \\
           &       &         & $1.56<|\eta|<1.81$& $0.420\pm0.003$ & $1.172\pm0.003$ & $0.420\pm0.003$ & $5.890\pm0.002$ & 46/9 \\
           &       &         & $1.81<|\eta|<2.37$& $0.419\pm0.003$ & $1.158\pm0.001$ & $0.420\pm0.002$ & $6.690\pm0.097$ & 52/8 \\
\hline
\end{tabular}}
\end{center}
\end{table*}

\begin{table*}[htbp]
\vspace{.0cm} \justifying\noindent {\small Table 4. Values of $T$,
$q$, $a_0$, $\sigma_0$, $\chi^2$, and ndof corresponding to the
dot-dashed curves in Figures 1(a)--8(a), which are fitted by the
revised Tsallis-like function when $\mu_i=0$ [Eq. (10)].}
\vspace{0mm}
\begin{center}
{\small
\begin{tabular}{lllllllll}
\hline \hline
Figure & Collab. & Energy & $\eta$ & $T$ (GeV) & $q$ & $a_0$ & $\sigma_0$ (mb) & $\chi^2$/ndof\\
\hline
Figure 1(a)& UA6   & 24.3 GeV& $-0.1<\eta<0.9$   & $0.120\pm0.003$  & $1.069\pm0.002$ & $0.205\pm0.001$ & $0.315\pm0.007$ & 10/5 \\
           & CCOR  & 62.4 GeV& $|\eta|<0.45$     & $0.113\pm0.002$  & $1.116\pm0.002$ & $0.232\pm0.002$ & $0.253\pm0.001$ & 17/4 \\
           & R806  & 63 GeV  & $|\eta|<0.2$      & $0.109\pm0.003$  & $1.113\pm0.001$ & $0.232\pm0.001$ & $0.194\pm0.003$ & 175/11 \\
           & R110  & 63 GeV  & $|\eta|<0.8$      & $0.115\pm0.002$  & $1.117\pm0.002$ & $0.230\pm0.001$ & $0.410\pm0.001$ & 5/3 \\
\hline
Figure 2(a)& PHENIX& 200 GeV & $|\eta|<0.35$     & $0.208\pm0.001$  & $1.146\pm0.002$ & $0.298\pm0.001$ & $0.079\pm0.002$ & 22/14 \\
           & NA24  & 300GeV  & $-0.65<\eta<0.52$ & $0.192\pm0.002$  & $1.132\pm0.002$ & $0.297\pm0.001$ & $0.055\pm0.003$ & 10/1 \\
           & CMS   & 2.76 TeV& $|\eta|<1.44$     & $0.231\pm0.002$  & $1.214\pm0.001$ & $0.348\pm0.001$ & $2.604\pm0.067$ & 4/1 \\
           & ALICE & 2.76 TeV& $|\eta|<0.9$      & $0.197\pm0.002$  & $1.182\pm0.003$ & $0.124\pm0.001$ & $8.132\pm0.102$ & 94/14 \\
\hline
Figure 3(a)& CMS   & 5.02 TeV& $|\eta|<1.44$     & $0.346\pm0.003$  & $1.203\pm0.002$ & $0.458\pm0.001$ & $0.580\pm0.004$ & 7/4 \\
           & CMS   & 7 TeV   & $|\eta|<1.45$     & $0.354\pm0.002$  & $1.206\pm0.001$ & $0.464\pm0.003$ & $0.660\pm0.003$ & 17/7 \\
           & ATLAS & 7 TeV   & $|\eta|<1.37$     & $0.351\pm0.002$  & $1.173\pm0.002$ & $0.466\pm0.003$ & $0.661\pm0.003$ & 112/9 \\
           & ALICE & 7 TeV   & $|\eta|<0.27$     & $0.361\pm0.002$  & $1.189\pm0.002$ & $0.474\pm0.002$ & $0.232\pm0.006$ & 45/5 \\
\hline
Figure 4(a)& ATLAS & 7 TeV   & $|\eta|<0.6$      & $0.359\pm0.003$  & $1.207\pm0.001$ & $0.476\pm0.002$ & $0.282\pm0.007$ & 18/12 \\
           &       &         & $0.6<|\eta|<1.37$ & $0.360\pm0.002$  & $1.202\pm0.003$ & $0.475\pm0.001$ & $0.462\pm0.004$ & 11/12 \\
           &       &         & $1.52<|\eta|<1.81$& $0.360\pm0.003$  & $1.202\pm0.004$ & $0.476\pm0.002$ & $0.070\pm0.003$ & 9/12 \\
           &       &         & $1.81<|\eta|<2.37$& $0.360\pm0.002$  & $1.202\pm0.003$ & $0.477\pm0.002$ & $0.242\pm0.012$ & 5/4 \\
\hline
Figure 5(a)& CMS   & 7 TeV   & $|\eta|<0.9$      & $0.360\pm0.002$  & $1.202\pm0.002$ & $0.478\pm0.001$ & $0.375\pm0.003$ & 6/11 \\
           &       &         & $0.9<|\eta|<1.44$ & $0.360\pm0.002$  & $1.202\pm0.004$ & $0.478\pm0.001$ & $0.218\pm0.004$ & 11/11 \\
           &       &         & $1.57<|\eta|<2.1$ & $0.360\pm0.002$  & $1.194\pm0.002$ & $0.478\pm0.001$ & $0.275\pm0.003$ & 46/11 \\
           &       &         & $2.1<|\eta|<2.5$  & $0.358\pm0.002$  & $1.198\pm0.001$ & $0.476\pm0.002$ & $0.150\pm0.004$ & 117/11 \\
\hline
Figure 6(a)& ATLAS & 8 TeV   & $|\eta|<0.6$      & $0.338\pm0.001$  & $1.192\pm0.003$ & $0.387\pm0.002$ & $3.302\pm0.014$ & 119/16 \\
           &       &         & $0.6<|\eta|<1.37$ & $0.356\pm0.004$  & $1.194\pm0.003$ & $0.344\pm0.001$ & $5.530\pm0.013$ & 273/16 \\
           &       &         & $1.56<|\eta|<1.81$& $0.335\pm0.004$  & $1.178\pm0.002$ & $0.348\pm0.001$ & $3.101\pm0.012$ & 118/14 \\
           &       &         & $1.81<|\eta|<2.37$& $0.340\pm0.002$  & $1.162\pm0.001$ & $0.348\pm0.003$ & $6.894\pm0.007$ & 640/14 \\
\hline
Figure 7(a)& CMS   & 13 TeV  & $|\eta|<0.8$      & $0.422\pm0.003$  & $1.192\pm0.001$ & $0.420\pm0.003$ & $2.260\pm0.005$ & 49/5 \\
           &       &         & $0.8<|\eta|<1.44$ & $0.424\pm0.003$  & $1.193\pm0.002$ & $0.424\pm0.003$ & $1.583\pm0.001$ & 38/5 \\
           &       &         & $1.57<|\eta|<2.1$ & $0.424\pm0.001$  & $1.166\pm0.002$ & $0.424\pm0.001$ & $4.290\pm0.014$ & 21/5 \\
           &       &         & $2.1<|\eta|<2.5$  & $0.420\pm0.005$  & $1.160\pm0.004$ & $0.419\pm0.003$ & $4.130\pm0.011$ & 4/4 \\
\hline
Figure 8(a)& ATLAS & 13 TeV  & $|\eta|<0.6$      & $0.423\pm0.001$  & $1.196\pm0.004$ & $0.423\pm0.002$ & $1.406\pm0.008$ & 112/10 \\
           &       &         & $0.6<|\eta|<1.37$ & $0.420\pm0.003$  & $1.195\pm0.001$ & $0.422\pm0.003$ & $2.340\pm0.003$ & 128/10 \\
           &       &         & $1.56<|\eta|<1.81$& $0.421\pm0.002$  & $1.172\pm0.003$ & $0.420\pm0.003$ & $5.390\pm0.096$ & 50/9 \\
           &       &         & $1.81<|\eta|<2.37$& $0.419\pm0.003$  & $1.158\pm0.002$ & $0.421\pm0.002$ & $6.790\pm0.095$ & 49/8 \\
\hline
\end{tabular}}
\end{center}
\end{table*}

\begin{figure*}[htbp]
\begin{center}
\includegraphics{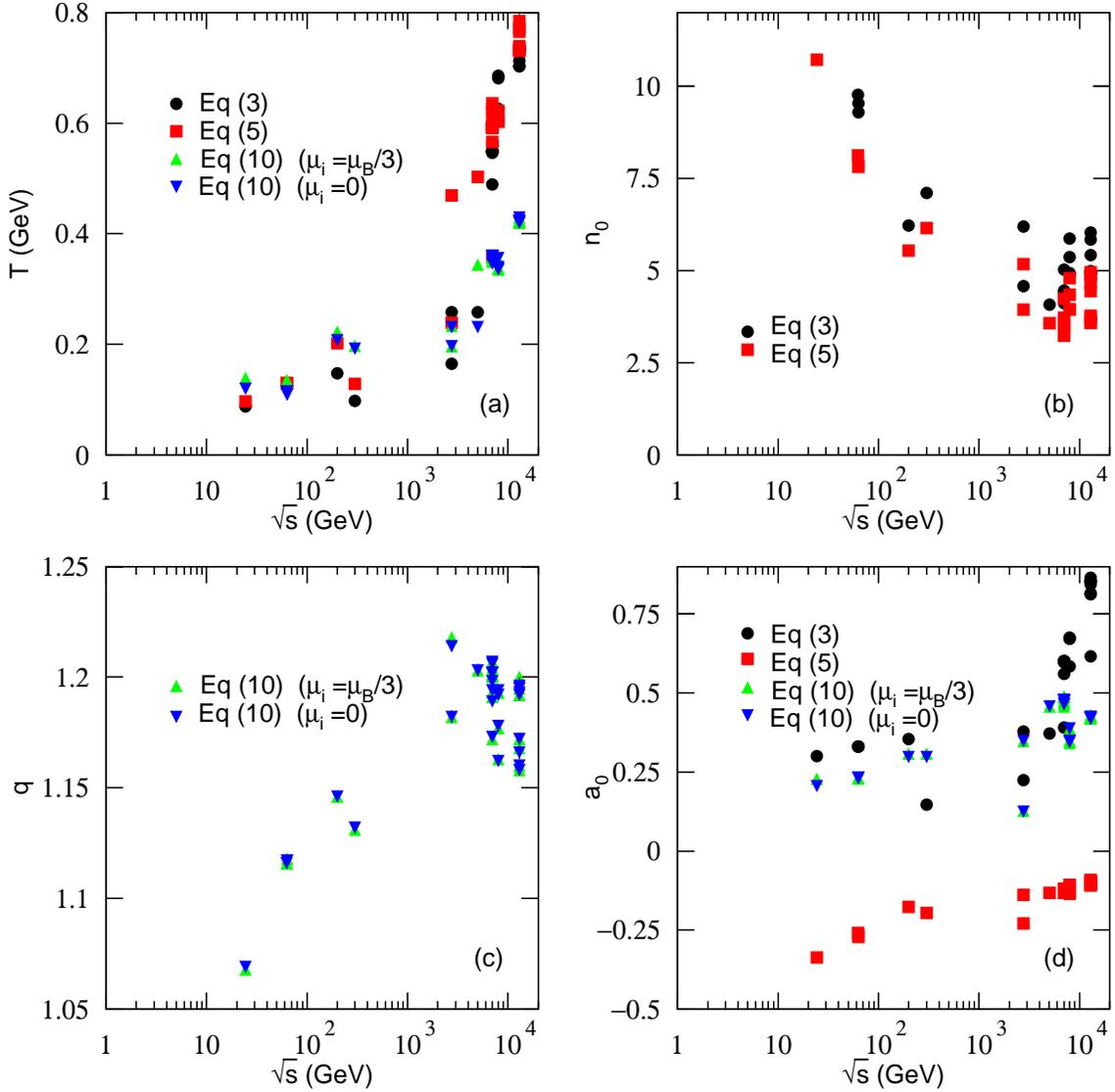}
\end{center}
\caption{\small Dependences of (a) $T$, (b) $n_0$, (c) $q$, and
(d) $a_0$ on $\sqrt{s}$. As marked in the panels, $T$ and $a_0$
are obtained from four cases, and $n_0$ and $q$ are obtained from
two cases.}
\end{figure*}

\begin{figure*}[htbp]
\begin{center}
\includegraphics{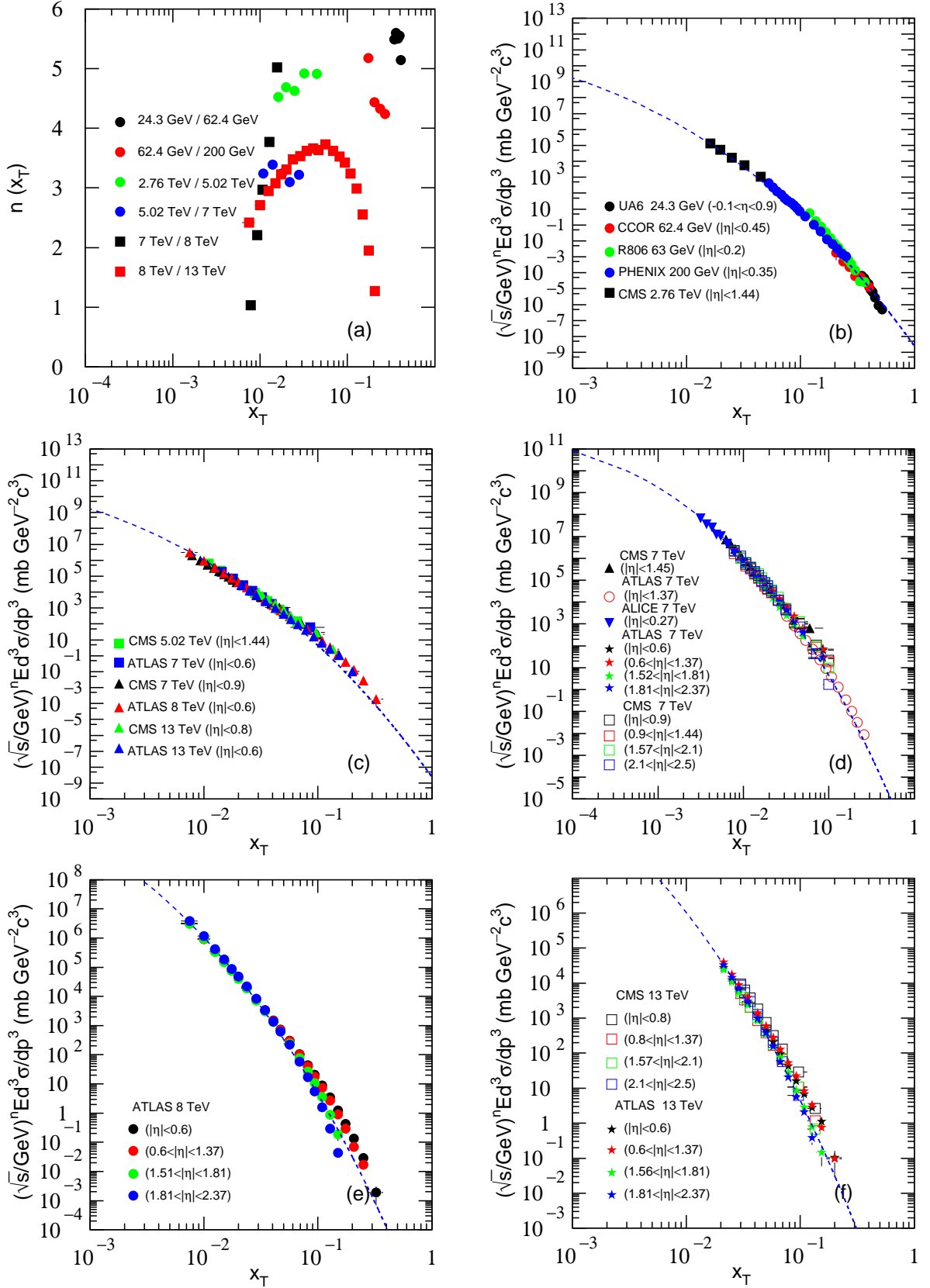}
\end{center}
\caption{\small (a) $n$ as the function of $x_T$. The symbols are
obtained from Eq. (13) due to the fittings. (b)--(f)
$(\sqrt{s}/{\rm GeV})^nE d^3\sigma/dp^3$ as the functions of $x_T$
at different energies with various $|\eta|$ marked in the panels.
The symbols are obtained from Eq. (15) due to the fittings, where
various $n$ are used as those listed in Table 5. The curves are
the result of fits by $G(x_T)$ presented in Eq. (16).}
\end{figure*}

\begin{table*}[htbp]
\vspace{.0cm} \justifying\noindent {\small Table 5. Summary of the
parameter $n$ for the spectra with the form of $x_T$ scaling in
p+p collisions. These $n$ are used for the symbols in Figures
10(b)--10(f) in which the curves are presented by Eq. (16).}
\vspace{0mm}
\begin{center}
{\small
\begin{tabular}{p{2.5cm}p{2.5cm}p{2.5cm}p{3cm}p{2.5cm}}
\hline \hline
Figure & Collab. & Energy & $\eta$ & $n$\\
\hline
Figure 10(b)& UA6   & 24.3 GeV& $-0.1<\eta<0.9$   & $2.87\pm0.06$\\
            & CCOR  & 62.4 GeV& $|\eta|<0.45$     & $3.42\pm0.04$\\
            & R806  & 63 GeV  & $|\eta|<0.2$      & $3.47\pm0.10$\\
            & PHENIX& 200 GeV & $|\eta|<0.35$     & $3.72\pm0.05$\\
            & CMS   & 2.76 TeV& $|\eta|<1.44$     & $3.16\pm0.04$\\
\hline
Figure 10(c)& CMS   & 5.02 TeV& $|\eta|<1.44$     & $3.22\pm0.05$\\
            &       & 7 TeV   & $|\eta|<0.9$      & $3.23\pm0.06$\\
            &       & 13 TeV  & $|\eta|<0.8$      & $3.21\pm0.04$\\
            & ATLAS & 7 TeV   & $|\eta|<0.6$      & $3.21\pm0.05$\\
            &       & 8 TeV   & $|\eta|<0.6$      & $3.20\pm0.03$\\
            &       & 13 TeV  & $|\eta|<0.6$      & $3.18\pm0.04$\\
\hline
Figure 10(d)& CMS   & 7 TeV   & $|\eta|<1.45$     & $3.26\pm0.06$\\
            &       &         & $|\eta|<0.9$      & $3.23\pm0.06$\\
            &       &         & $0.9<|\eta|<1.44$ & $3.22\pm0.05$\\
            &       &         & $1.57<|\eta|<2.1$ & $3.26\pm0.02$\\
            &       &         & $2.1<|\eta|<2.5$  & $3.25\pm0.03$\\
            & ALICE & 7 TeV   & $|\eta|<0.27$     & $3.21\pm0.05$\\
            & ATLAS & 7 TeV   & $|\eta|<1.37$     & $3.02\pm0.05$\\
            &       &         & $|\eta|<0.6$      & $3.21\pm0.05$\\
            &       &         & $0.6<|\eta|<1.37$ & $3.21\pm0.03$\\
            &       &         & $1.52<|\eta|<1.81$& $3.22\pm0.11$\\
            &       &         & $1.81<|\eta|<2.37$& $3.22\pm0.08$\\
\hline
Figure 10(e)& ATLAS & 8 TeV   & $|\eta|<0.6$      & $3.20\pm0.03$\\
            &       &         & $0.6<|\eta|<1.37$ & $3.17\pm0.02$\\
            &       &         & $1.36<|\eta|<1.81$& $3.29\pm0.02$\\
            &       &         & $1.81<|\eta|<2.37$& $3.23\pm0.03$\\
\hline
Figure 10(f)& CMS   & 13 TeV  & $|\eta|<0.8$      & $3.26\pm0.04$\\
            &       &         & $0.8<|\eta|<1.44$ & $3.20\pm0.05$\\
            &       &         & $1.57<|\eta|<2.1$ & $3.22\pm0.04$\\
            &       &         & $2.1<|\eta|<2.5$  & $3.29\pm0.07$\\
            & ATLAS & 13 TeV  & $|\eta|<0.6$      & $3.18\pm0.04$\\
            &       &         & $0.6<|\eta|<1.37$ & $3.15\pm0.02$\\
            &       &         & $1.56<|\eta|<1.81$& $3.26\pm0.03$\\
            &       &         & $1.81<|\eta|<2.37$& $3.22\pm0.03$\\
\hline
\end{tabular}}
\end{center}
\end{table*}

In the fits, we have used $m_0=0$ for direct photon and
$m_{0i}=0.31$ GeV/$c^2$ for participant parton in the two-body
process, where the heavier partons are not considered due to their
very small production probability. The least square method is used
to determine the values of free parameters and normalization
constant, and the statistical simulation method is used to
determine the errors. The free parameters ($T$, $n_0$, and $a_0$),
the normalization constant ($\sigma_0$), $\chi^2$, and the number
of degrees of freedom (ndof) obtained from the fits by Eqs. (3)
and (5) are listed in Tables 1 and 2 respectively. The values of
$T$, $q$, $a_0$, $\sigma_0$, $\chi^2$, and ndof obtained from the
fits by Eq. (10) with $\mu_i=\mu_B/3$ and $\mu_i=0$ are listed in
Tables 3 and 4 respectively.

In order to compare the degree of deviation between the fitting
results and the data, Figures 1(b)--1(e) show the ratios of data
to fit obtained from four cases: Eqs. (3), (5), (10) with
$\mu_i=\mu_B/3$, and (10) with $\mu_i=0$, orderly. One can see
that the fit results reproduce satisfactorily the data in the
whole $p_T$ range. As seen from Figure 1(a) and corresponding
results of Tables 1--4, the fits in four cases are in satisfactory
agreement with the experimental data. Although the data of direct
photons generated in p+p collisions measured by the UA6, R108,
R806, and R110 Collaborations at 24.3, 62.4, and 63
GeV~\cite{61,62,63,64} can be fitted by Eqs. (3), (5), and (10),
the mid-$\eta$ ranges used in different experiments are not the
same exactly. These differences do not affect significantly the
results due to the fact that we have used the results per $\eta$
unit.

It seems that we do not need to use so many fitting functions,
because Eq. (3) fits well enough the experimental data. However,
the fits using Eqs. (5) and (10) can reveal more quantities due to
different physical ideas used. In particular, the results fitted
by various functions can be compared with each other. So it is
useful to use different functions to fit the data.

Similar to Figure 1(a), Figures 2(a)--8(a) show the invariant
cross-section of direct photons produced in two-body processes in
p+p collisions at higher energies. The curves have the same
meanings as Figure 1(a), and the parameters are listed in Tables
1--4.

In Figure 2(a), the points (symbols) in different colors represent
orderly the data measured by the PHENIX Collaboration at 200 GeV
with $|\eta|<0.35$~\cite{65}, the NA24 Collaboration at 300 GeV
with $-0.65<\eta<0.52$~\cite{66}, the CMS Collaboration at 2.76
TeV with $|\eta|<0.9$~\cite{67}, and the ALICE Collaboration at
2.76 TeV with $|\eta|<1.44$~\cite{68}, with different re-scaling
factors ($10^6$, $10^5$, $10^{-3}$, and $10^{-7}$) marked in the
panel.

In Figure 3(a), the data correspond orderly to the results
measured by the CMS Collaboration at 5.02 TeV with
$|\eta|<1.44$~\cite{69} and at 7 TeV with $|\eta|<1.45$~\cite{70},
as well as the results measured by the ATLAS Collaboration at 7
TeV with $|\eta|<1.37$~\cite{71} and by the ALICE Collaboration at
7 TeV with $|\eta|<0.27$~\cite{73}, with different re-scaling
factors ($10^6$, $10^3$, $10^0$, and $10^{-3}$) marked in the
panel, where the re-scaling factor $10^0$ is not marked.

The symbols in Figure 4(a) represent orderly the data measured by
the ATLAS Collaboration at 7 TeV with $|\eta|<0.6$,
$0.6<|\eta|<1.37$, $1.52<|\eta|<1.81$, and
$1.81<|\eta|<2.37$~\cite{72,74}. The symbols in Figure 5(a)
represent orderly the data measured by the CMS Collaboration at 7
TeV with $|\eta|<0.9$, $0.9<|\eta|<1.44$, $1.57<|\eta|<2.1$, and
$2.1<|\eta|<2.5$~\cite{75}. Different re-scaling factors ($10^2$,
$10^0$, $10^{-2}$, and $10^{-4}$) are used for clarity.

In Figure 6(a), the symbols represent orderly the data measured by
the ATLAS Collaboration at 8 TeV with $|\eta|<0.6$,
$0.6<|\eta|<1.37$, $1.56<|\eta|<1.81$, and
$1.81<|\eta|<2.37$~\cite{76}. In Figure 7(a), the symbols
represent orderly the data measured by the CMS Collaboration at 13
TeV with $|\eta|<0.8$, $0.8<|\eta|<1.44$, $1.57<|\eta|<2.1$, and
$2.1<|\eta|<2.5$~\cite{77}. In Figure 8(a), the symbols represent
orderly the data measured by the ATLAS Collaboration at 13 TeV
with $|\eta|<0.6$, $0.6<|\eta|<1.37$, $1.56<|\eta|<1.81$, and
$1.81<|\eta|<2.37$~\cite{78}. In the three figures, the re-scaling
factors are orderly $10^4$, $10^2$, $10^0$, and $10^{-2}$.

The ratios of data to fit in Figures 2--8 are the same as in
Figure 1. In some cases, the deviations between the fits and data
are significant, which means that the fits are approximate. In
most cases, the fits are satisfactory and acceptable. At least,
the trends of data can be revealed from the fits by Eqs. (3), (5),
and (10). From the values of $\chi^2$ listed in Tables 1--4, one
can see that the fitting qualities of the three equations in four
cases are nearly the same. It is hard to judge which one is
better. In our opinion, the method using the convolution of two
TP-like functions is more convenient to be extended to the
convolution of three or more functions and to be explained by the
underlying multi-parton process, even performing the convolution
by other functions.

Concerning the fit quality, from Tables 1--4 one can see that the
values for $\chi^2/$ndof are systematically higher than one. On
the other hand, fits using Tsallis-like distribution on particle
level present consistently $\chi^2/{\rm ndof}<1$ in case of
charged particle production in p+p collisions~\cite{27}. The
reason is that the present work deals with the data sample for
direct (prompt) photons which have lower statistics and wider
$p_T$ range than those for charged particles in experiments.

\subsection{Trends of parameters and discussion on methodology}

We now study the variation trend of the extracted free parameters.
Figures 9(a)--9(d) show the dependences of effective temperature
$T$, power index $n_0$, entropy index $q$, and correction index
$a_0$ on collision energy $\sqrt{s}$, respectively. The symbols
represent the values of parameters listed in Tables 1--4 and
extracted from Figures 1--8 by using Eqs. (3), (5), and (10). One
can see that with the increase of $\sqrt{s}$, the parameters $T$,
$q$, and $a_0$ increase, and the parameter $n_0$ decreases.

The results from Figure 9 is understandable. At higher energy, the
collision system stays at higher excitation state with lesser
probability of equilibrium. The parameters $T$ and $q$ ($n_0$) are
naturally larger (smaller). In addition, the spectra of $p_T$ are
wider at higher energy. This causes a valley in the spectra at
very low $p_T$, which needs a larger $a_0$ to fit it. In some
references (for instance~\cite{27,80a}) it is proposed a QCD
inspired model for the energy dependence of parameters $T$ and $q$
concerning charged particle production. In general, a
$\ln\sqrt{s}$ dependence is presented. From Figure 9, it seems
this is not the case (it is steeper) for direct (prompt) photons.
The reason is that, compared with charged particles, direct
(prompt) photons are emitted earlier in the source with higher
excitation degree. In addition, resonance decay and medium effect
affect charged particle production more significantly. These
factors cause different behaviors for two types of particles.

The effective temperature is fitted in particle level and in
parton level. Both of them contain the contributions of thermal
motion and transverse flow effect (if available in p+p
collisions). The latter should be excluded from the contributions
to extract the real temperature. As a non-real temperature, the
effective temperature is model-dependent. So, the effective
temperatures obtained from different functions may be different.
In addition, direct (prompt) photons are emitted earlier than
light charged particles in general. This is also why the effective
temperature obtained from this work is quite higher than ones
typically obtained in Tsallis-like fits to light charged particle
production~\cite{48a,80b}.

In most cases, Figure 9 shows that the parameters from different
equations are similar to each other. However, in Figure 9(d),
although the same symbol $a_0$ is used, the values may be
different in Eqs. (3) and (5) due to Eq. (3) having the item
$\big(p_T\sqrt{p_T^2+m_0^2}\,\big)^{a_0}$ and Eq. (5) having the
product of $\big(p_{t1}\sqrt{p_{t1}^2+m_{01}^2}\,\big)^{a_0}$ and
$\big[(p_T-p_{t1})\sqrt{(p_T-p_{t1})^2+m_{02}^2}\,\big]^{a_0}$.
The influence of $a_0$ in Eq. (5) is double. This renders the
difference in $a_0$ from Eqs. (3) and (5). This difference does
not happen in Eq. (10) due to another calculation.

Figure 10(a) shows $n$ as the function of $x_T$, and Figures
10(b)--10(f) shows $(\sqrt{s}/{\rm GeV})^nE d^3\sigma/dp^3$ as the
functions of $x_T$ at different energies with various $|\eta|$
marked in the panels. From Figure 10(a), one cannot see a
monotonous dependence in terms of the function $n$. From Figures
10(b)--10(f), in terms of the function $(\sqrt{s}/{\rm GeV})^nE
d^3\sigma/dp^3$, one can see that the data of different $p_T$
follow a similar curve. Selecting appropriate $n$ at given energy,
an energy independent $x_T$ scaling can be obtained. The scaling
function in Figures 10(b)--10(f) can be empirically described by
\begin{align}
G(x_T)=(\sqrt{s}/{\rm GeV})^nE\frac{d^3\sigma}{dp^3}=a\times
x_T^{b \times x_T^{-c}},
\end{align}
where $a=(2.4403\pm0.4181)\times 10^{-8}$, $b=-10.0129\pm0.1260$,
and $c=-0.0837\pm0.0017$ with various $n$ listed in Table 5. As a
global fit for Figures 10(b)--10(f), we have obtained $\chi^2/{\rm
ndof}=19/350$. In the function, the cross-section, energy, and
momentum are in the units of mb, GeV, and GeV/$c$, respectively.
From Figures 10(b)--10(f), one can see that Eq. (16) is
approximately an anti-correlation in double logarithmic
representations.

The form of the function $(\sqrt{s}/{\rm GeV})^nE d^3\sigma/dp^3=
(\sqrt{s}/{\rm GeV})^n (1/2\pi p_T)d^2\sigma/dp_Tdy$ in the
dependence on the variable $x_T=2p_T/\sqrt{s}$ can be empirically
constructed~\cite{80}. The concrete form obtained in the present
work is a simple expression with three parameters. This scaling
law renders that there is common similarity and universality in
many aspects in high energy
collisions~\cite{81,81a,81b,82,83,84,85,86}. The underlying
physics reason is that participant quarks or partons from the
collision system take part in the considered scattering process,
which is related to the parton saturation and higher-order QCD
contribution from $qg$ and $q\bar q$ channels.

The functions which are used to fit the spectra in the present
work are not sole ones. Although we have not presented the related
results, Eq. (7) is also suitable to fit the spectra due to small
difference from Eq. (3). As mentioned in the second section,
$n_0=1/(q-1)$ and $q$ is close to 1. This causes Eq. (7) to be
similar to Eq. (3). However, the difference between Eqs. (5) and
(10) is large. Their results are only similar in the available
data region in the $p_T$ spectra.

We would like to point out that Eqs. (5) and (10) are two special
cases from two participant partons. If the vectors ${\bm p_{t1}}$
and ${\bm p_{t2}}$ are parallel, we have $p_T=p_{t1}+p_{t2}$ and
Eq. (5). If the vectors ${\bm p_{t1}}$ and ${\bm p_{t2}}$ are
perpendicular, we have $p_T=\sqrt{p_{t1}^2+p_{t2}^2}$ and Eq.
(10). If the vectors ${\bm p_{t1}}$ and ${\bm p_{t2}}$ form an
arbitrary angle, the analytical expression of $p_T$ is not
available for us. Instead, we may use the Monte Carlo calculation
to give many concrete $p_T$ by using the relation
$p_T=\sqrt{(p_{t1}\cos\phi_1+p_{t2}\cos\phi_2)^2+(p_{t1}\sin\phi_1+p_{t2}\sin\phi_2)^2}$,
where $\phi_1$ and $\phi_2$ are the azimuth angles of the vectors
${\bm p_{t1}}$ and ${\bm p_{t2}}$. The values of $\phi_1$ and
$\phi_2$ can be evenly distributed in $[0,2\pi]$, and the values
of $p_{t1}$ and $p_{t2}$ obey Eq. (4) or (8). After repeated
calculations for many times, the distribution of $p_T$ can be
obtained by statistics. Our exploratory research shows that the
Monte Carlo calculation is suitable.

If the participants are three or more partons, one may extend Eqs.
(5) and (10) in principal. In fact, one may consider the
contributions of the first two partons, and combine the
contributions with the third one. Then, one may combine the
contributions of the first three partons with the fourth one, and
so on. In the case of the number of participants being large, the
analytical expression becomes complex and inconvenient. Instead,
one may use the Monte Carlo calculation for the two special cases
and the general case. What one does is just adding items in the
expressions of $p_T$ for the three or more participant partons.

Before summary and conclusions, we would like to collect a few
other physical approaches describing the spectra of direct
(prompt) photons. For examples, (i) QCD color dipole
picture~\cite{97a,97b}, in which the low-$p_T$ region is described
by the parton saturation effects and pQCD evolution is reproduced
accurately. The $x_T$ scaling can be also addressed in this
formalism. (ii) High energy or $k_T$-factorization
approach~\cite{97c,97d}, which is based on off-shell matrix
elements to compute the sub-process cross section. Building blocks
include unintegrated parton distribution functions (uPDFs). (iii)
Next-to-leading-order (NLO), next-to-next-to-leading-order (NNLO),
and next-to-next-to-next-to-leading-order (NNNLO) QCD
calculations~\cite{97e,97f,97g}, in which the well known pQCD
approach usually implemented in the Monte Carlo simulations. (iv)
Color Glass Condensate effective field theory (CGC
EFT)~\cite{97h,97i,97j,97k,97l,97m}, in which the dipole forward
scattering amplitude is given in terms of fundamental lightlike
Wilson lines. It is known at NLO accuracy level. The present paper
is consistent with these approaches.

In addition, we would like to emphasize the significance of the
present work. We have used two types of parametrization which come
from the statistical method to fit the direct (prompt) photon
spectra as measured in p+p collisions in a wide range of colliding
energy and found a scaling form to describe the spectra. Although
the direct (prompt) photon spectra with large $p_T$ at collider
energies can be described by pQCD and similar theoretical
treatments, the present work proposes an alternative method, i.e.
the statistical method that was developed for charged particle
production in high-energy heavy ion collisions, to describe the
direct (prompt) photon spectra in p+p collisions. These fits
conform that the underlying participants in p+p collisions are
partons, but not protons. Participants are partons, which is the
source of the common similarity and
universality~\cite{81,81a,81b,82,83,84,85,86}.

{\section{Summary and conclusions}}

We summarize here our main observations and conclusions.

(a) In this paper, we have analyzed the transverse momentum
spectrum of direct photons generated at mid-rapidity in p+p
collisions over an energy range from 24.3 GeV to 13 TeV. We have
conducted the research through two methods: (i) The TP-like
function at the particle level and the convolution of two TP-like
functions at the parton level; (ii) The revised Tsallis-like
function at the particle level and the root-sum-of-squares of two
revised Tsallis-like functions at the parton level. The calculated
results can be used to fit the experimental data measured by
international collaborations.

(b) In the first method, the $p_T$ value of direct photon is
directly regarded as both the subjecting to the TP-like function
and the equaling to the sum of transverse momenta of two
participant partons. The contribution of each participant parton
to the $p_T$ spectrum of particles is also the TP-like function.
In the second method, we regard $p_T$ of direct photon as both the
subjecting to the revised Tsallis-like function and the equaling
to the root-sum-of-squares of transverse momenta of two
participant partons. The transverse momentum of partons also obeys
the revised Tsallis-like function. In the two methods, we describe
the $p_T$ spectrum not only at the particle level, but also at the
parton level in the framework of multi-source thermal model. The
fitting results of the two methods are basically consistent with
the experimental data. Although it is hard to judge which case is
more suitable due to the incomplete $p_T$ range in data, the idea
at the parton level in the first method is more convenient to be
extended to the scene of multiple participant partons.

(c) The two methods correspond to two special cases of the
relation between the transverse momenta of two partons. In the
first method, the transverse momenta of two partons are parallel.
In the second method, the transverse momenta of two partons are
perpendicular. In the general case, the transverse momenta of two
partons may have any azimuthal angles in $[0,2\pi]$. The Monte
Carlo calculation can be used to obtain the $p_T$ value of
particle. The main parameters are the effective temperature $T$,
power index $n_0$ (or entropy index $q$), and correction index
$a_0$. Although the absolute values of parameters are different
due to different cases, the observed trends of parameters are
similar. This work shows that $T$, $q$, and $a_0$ increase and
$n_0$ decreases with the increase of collision energy.

(d) Based on the analyses of $p_T$ spectra of direct photons, an
energy independent scaling, i.e. the $x_T$ scaling, is obtained to
show approximately anti-correlation in double logarithmic
representations. Not only for the spectra in low-$p_T$ region, but
also for the spectra in high-$p_T$ region, the data follow
approximately the function $(\sqrt{s}/{\rm GeV})^nE
d^3\sigma/dp^3$ of the variable $x_T=2p_T/\sqrt{s}$. Suitable $n$
may show an energy independent scaling which is expressed by
$(\sqrt{s}/{\rm GeV})^nE d^3\sigma/dp^3=a\times x_T^{b \times
x_T^{-c}}$. This scaling law renders that there are many
similarities and universality in some distributions in various
collisions at high energies. The underlying physics reason is that
the participant partons, but not nucleons, from the collision
system take part in the two-body scattering process, which
involves the parton saturation and higher-order QCD contribution
from $qg$ and $q\bar q$ channels.
\\
\\
\\
{\bf Acknowledgments} The work of Q.Z. and F.H.L. was supported by
the National Natural Science Foundation of China under Grant Nos.
12147215, 12047571, 11575103, and 11947418, the Shanxi Provincial
Natural Science Foundation under Grant No. 201901D111043, and the
Fund for Shanxi ``1331 Project" Key Subjects Construction. The
work of Y.Q.G. was supported by the Scientific and Technological
Innovation Programs of Higher Education Institutions in Shanxi
(STIP) under Grant No. 2019L0629. The work of Kh.K.O. was
supported by the Ministry of Innovative Development of the
Republic of Uzbekistan within the fundamental project No.
F3-20200929146 on analysis of open data on heavy-ion collisions at
RHIC and LHC.
\\
\\
{\bf Author Contributions} All authors listed have made a
substantial, direct, and intellectual contribution to the work and
approved it for publication.
\\
\\
{\bf Data Availability Statement} This manuscript has no
associated data or the data will not be deposited. [Authors'
comment: The data used to support the findings of this study are
included within the article and are cited at relevant places
within the text as references.]
\\
\\
{\bf Ethical Approval} The authors declare that they are in
compliance with ethical standards regarding the content of this
paper.
\\
\\
{\bf Disclosure} The funding agencies have no role in the design
of the study; in the collection, analysis, or interpretation of
the data; in the writing of the manuscript, or in the decision to
publish the results.
\\
\\
{\bf Conflict of Interest} The authors declare that there are no
conflicts of interest regarding the publication of this paper.
\\
\\

\end{document}